# The Effects of Cobalt Doping on the Skyrmion Hosting Material $Cu_2OSeO_3$

M. Vás[1-3], A. J. Ferguson[4], H. E. Maynard-Casely[5], C. Ulrich[6*], E. P. Gilbert[5], S. Yick[1-2*], T. Söhnel[1-2*].

1. School of Chemical Sciences, The University of Auckland, Auckland, New Zealand

2. MacDiarmid Institute for Advanced Materials and Nanotechnology, Wellington, New Zealand

3. Australian Institute of Nuclear Science and Engineering, Lucas Heights, New South Wales, Australia

4. Department of Physics, University of Fribourg, Fribourg, Switzerland

5. Australian Centre for Neutron Scattering, ANSTO, Lucas Heights, New South Wales, Australia

6. School of Physics, University of New South Wales, Sydney, New South Wales, Australia

*Corresponding authors: c.ulrich@unsw.edu.au, samuel.yick@auckland.ac.nz, t.soehnel.@auckland.ac.nz

## Abstract

$Cu_2OSeO_3$ is a non-centrosymmetric multiferroic helimagnet that stabilises magnetic skyrmion lattices via competing exchange interactions and thermal fluctuation. We examine the role of cobalt substitution on its structural and magnetic ordering. Synchrotron X-ray and neutron diffraction confirm successful $Co^{2+}$ doping into the crystal structure. Magnetisation and small-angle neutron scattering measurements show that this substitution perturbs the exchange pathways, leading to enhanced critical fields and lower ordering temperatures. Notably, Co-doping drives a systematic expansion and shift of the skyrmion stability pocket. These results demonstrate that targeted incorporation of magnetic ions provides a controlled route to tune exchange competition and skyrmion phase stability in $Cu_2OSeO_3$, offering new insights into the manipulation of chiral spin textures in helimagnets.

# I. Introduction

Magnetic skyrmions are topologically protected quasiparticles, as they are characterised by a non-zero topological integer.[1,2] They arrange themselves in a vortex-like magnetic spin structure that appears on the nanoscale (10 – 100 nm), typically in a hexagonal lattice in magnetic materials.[3] The formation of magnetic skyrmions is dependent on the competing symmetric exchange interactions (SEI) and Dzyaloshinskii-Moriya interactions (DMI) that arise due to the non-centrosymmetric arrangement of the material's crystal structure.[4] These magnetic skyrmions are only stable in a restricted window of temperature and magnetic fields, known as the skyrmion pocket. Skyrmions were first proposed in 1962 as a concept to understand stable localised field configurations of hadrons by using unified field theory.[5,6] However, in 2008, they were found to experimentally describe a magnetic feature observed in MnSi via small-angle neutron-scattering (SANS).[7] Since then, magnetic skyrmions have been observed in other $P2_13$ space group (B20 class) materials such as FeGe and $Fe_{1-x}Co_xSi$.[8–13] Magnetic skyrmions have also been observed in real space using Lorentz transmission electron microscopy (LTEM), which shows the formation and destruction of these spin structures under applied magnetic fields, currents and temperatures.[9,14,15] Despite the different chemical compositions of these materials, they all crystallise in the cubic $P2_13$ space group and exhibit similar magnetic phases.[2] This suggests that the crystal structure influences the capability of the material to host magnetic skyrmions.

The $Cu_2OSeO_3$ material system has gained significant attention as it is the only multiferroic insulator known to host magnetic Bloch-type skyrmions. $Cu_2OSeO_3$ also crystallises in the non-centrosymmetric cubic $P2_13$ space group, but is a complex metal oxide rather than an intermetallic compound, which allows for easier doping of the crystal structure through changing the stoichiometry of the metal oxides.[2] The unit cell contains two crystallographic sites that the $Cu^{2+}$ ions occupy, with their Wyckoff positions being 4a for the Cu1 site and 12b for the Cu2 site. All $Cu^{2+}$ ions are coordinated to five oxygens, forming $CuO_5$ polyhedra.[16] The Cu1 polyhedra are arranged in a distorted trigonal bipyramidal coordination, while the Cu2 polyhedra are arranged in a distorted square pyramidal coordination. The $Cu^{2+}$ sites form a pyrochlore-type network of distorted tetrahedra comprising one Cu1 site and three Cu2 sites, which is responsible for the material's magnetic properties. The three Cu2 sites have their spins aligned parallel, while the Cu1 site is aligned antiparallel, resulting in a three-spin up one-spin down ferrimagnetic structure when below $T_C \sim 58$ K, illustrated in Fig. 1.[16] The exchange integrals in the system lead to them being positive between two Cu2 sites, which have ferromagnetic (FM) alignment of spins, and negative between Cu1 and Cu2 sites, which have antiferromagnetic (AFM) alignment.[17]

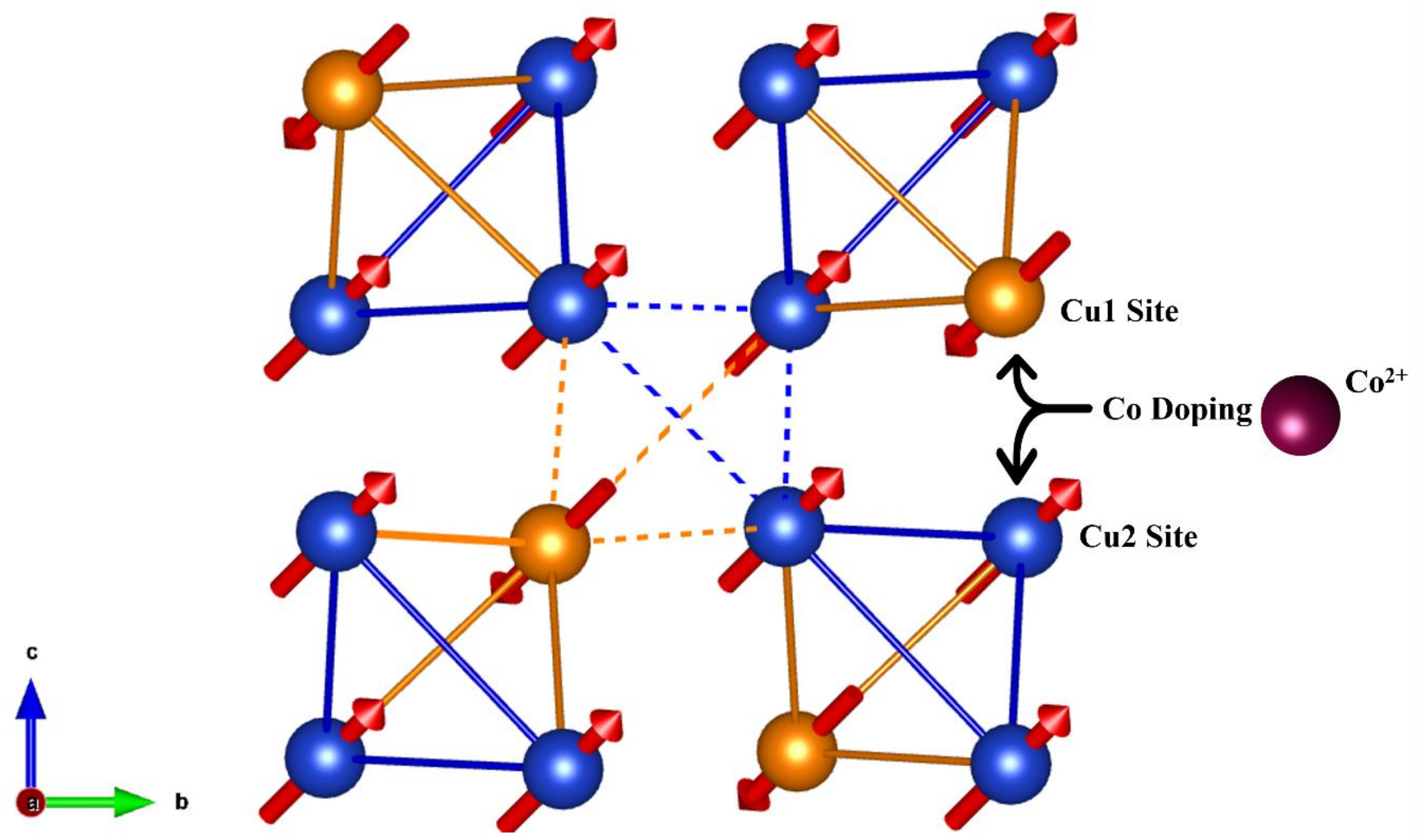

**FIG. 1.** The magnetic unit cell of $Cu_2OSeO_3$ shows the structure of the magnetic $Cu^{2+}$ ions, with the Cu1 and Cu2 sites being represented by the orange and blue atoms, respectively. Structurally, cobalt can dope into either copper site.

This system has been further investigated by substituting both the non-magnetic site (Se) and the magnetic site (Cu). Substituting atoms of different sizes into the copper and selenium sites impacts the magnetic interactions, which, in turn, influences the formation of magnetic skyrmions in this material.

Furthermore, substituting magnetic ions of different moments into the $Cu^{2+}$ site can directly influence magnetic exchange interactions. Understanding the role that distortion to the magnetic network plays in the formation and stability of topological spin structures will be vital for designing materials for spintronics applications.

Previously, researchers have investigated the effects of doping nickel, zinc, and silver into the copper sites at various doping levels in both polycrystalline and single crystal samples.[18–23] Through magnetic measurements, the doping has been shown to affect the skyrmion pocket by stabilising and widening it towards lower temperatures. The effects of nickel and zinc doping have also been previously studied using SANS, which showed a reduction in the critical temperatures for both dopants; it was found that doping with another magnetic ion, $Ni^{2+}$, enhanced the DMI in the system, increasing the temperature stability of the skyrmions, while doping with a non-magnetic ion, $Zn^{2+}$, weakened the DMI and therefore decreased the temperature stability of the skyrmions.[23] The effect of Ag doping seemed to show enhancement of the helical stability with respect to the field, along with a decrease in the saturation field.[22]

Additionally, the effects of doping the non-magnetic site have been studied by replacing selenium with tellurium in both single-crystal and polycrystalline samples.[14,24,25] In Te-doped single crystals, we showed with LTEM that the presence of dopant increased the skyrmion size and the helical modulation period.[14] It was originally claimed in polycrystalline samples that increasing Te-doping decreased the temperature range of the skyrmion pocket, along with reducing the temperature of the helical to paramagnetic transition using magnetometry measurements.[24] However, we were able to show with magnetometry and small angle neutron scattering that with increasing Te-doping, the temperature range of the skyrmion pocket expanded along with a shift to lower temperatures.[25]

Despite the plethora of work on chemically modifying $Cu_2OSeO_3$, to date no report of cobalt doping into the crystal structure has been published. The nature of $Co^{2+}$ ions makes their introduction into the magnetic site much more interesting, as high-spin $Co^{2+}$ has a slightly larger ionic radius than $Cu^{2+}$, unlike $Ni^{2+}$; thus, we expect an expansion of the crystal lattice upon doping. However, the more pronounced difference lies in their magnetic moments: 3.87 $\mu$B for $Co^{2+}$ and 1.73 $\mu$B for $Cu^{2+}$.[26] Thus, depending on the site, the exchange interaction and the resulting magnetic structures of the system can be drastically affected. In this work, we present a detailed study of doping $Co^{2+}$ into the $Cu^{2+}$ sites and present changes to the magnetic interactions with both DC magnetometry and SANS measurements.

## II. Experimental Methods

### A. Sample Preparation

Polycrystalline samples of $(Cu_{1-x}Co_x)_2OSeO_3$ ($0 \leq x \leq 0.2$) were prepared by solid-state sintering. Stoichiometric amounts of high-purity CuO (99.5% - Alfa Aesar), CoO (99.99% - Sigma Aldrich) and $SeO_2$ (99.99% - Sigma Aldrich) powders were mixed into homogenous mixtures before being sealed in evacuated quartz tubes and placed in a tube furnace. The sealed ampules were heated to 610 °C over 2 hours, held at the sintering temperature for 48 hours, before being slowly cooled to room temperature over 6 hours. Sample phase purity was confirmed with Rietveld refinements of synchrotron pXRD data.

### B. Modelling

The structural properties of the polycrystalline samples were determined from the room-temperature synchrotron pXRD and room temperature high-resolution NPD data using Rietveld refinements performed with the FullProf software.[27–29] The model crystal structure unit cell shown in Fig. 1 was modelled in VESTA 3.[30]

### C. Powder X-ray Diffraction

The pXRD data were collected on the Powder Diffraction Beamline at the Australian Synchrotron, ANSTO (Melbourne, Australia). For all measurements, each polycrystalline sample was loaded into a glass capillary with an outer diameter of 0.3 mm before being measured by the automatic sample-changing robot. Data were collected at room temperature using 21 keV X-rays with a wavelength of 0.59053(1) Å, in two scans of 150 seconds spliced together to account for gaps along 2θ in the Mythen-II detector.

## D. Neutron Powder Diffraction

The NPD data were collected on the ECHIDNA instrument, a high-resolution neutron powder diffractometer at the Australian Centre for Neutron Scattering (ACNS), ANSTO (Lucas Heights, Australia).[31] For all measurements, approximately 2 g of each polycrystalline sample was filled into 9 mm vanadium canisters before being sealed and loaded into the ECHIDNA sample changer robot to carry out the room temperature measurements. A 2.44 Å wavelength neutron beam was used, with a [331] Ge single-crystal monochromator, to scan the samples for 8 hours each.

## E. Elemental Analysis

Elemental analysis of the polycrystalline samples was performed using EDS and XAS, and Rietveld refinements of the NPD data. The EDS data were collected on a ZEISS GeminiSEM 460 microscope at the Centre for Electron Microscopy and Materials Science (CEMMS), University of Auckland (Auckland, New Zealand). All polycrystalline samples were loaded onto SEM stubs and imaged in backscattered electron (BSE) mode using a beam voltage of 20 keV, probe current of 1 nA, at a working distance of approximately 9.3 mm. The EDS spectra were collected with the acquisition mode set to auto using a BSE detector and processed with the software, Aztec, by Oxford Instruments. Each sample had five EDS spectra collected from different areas to obtain an average elemental analysis result. The XAS data at the cobalt *K*-edge were collected on the Medium Energy X-ray Absorption Spectroscopy (MEX-1) beamline at the Australian Synchrotron, ANSTO (Melbourne, Australia). The X-ray absorption near-edge structure (XANES) data for the cobalt *K*-edge, located at 7.709 keV, were obtained in transmission mode. 18 data points with 0.01 keV step sizes between 7.509 – 7.689 keV, 350 data points with 0.0002 keV step sizes between 7.689 – 7.759 keV and 89 data points with 0.005 keV step sizes between 7.759 – 8.200 keV, all scanned for one second. The interpretation and fitting of the XAS data were performed using the ATHENA software suite.[32]

### F. Magnetometry Measurements

Magnetometry was performed on a SQUID magnetometer (Quantum Design MPMS3) with the DC VSM option. Approximately 40 mg of each of the polycrystalline samples was loaded into powder capsules and attached to a brass sample holder. Magnetisation-temperature measurements were performed over the range of 2 – 80 K at zero applied field after being ZFC cooled down to base temperature. The magnet was reset by quenching it prior to sample cooling to eliminate any remnant field. The samples were measured upon continuous heating from 2 – 40 K at 1 K/min and from 41 – 80 K at 0.3 K/min. For the M-H sweeps, prior to each measurement, samples were first cooled from 80 to 2 K and warmed to the desired temperature while at zero field, before taking the field measurement. Finally, the sample was heated to 80 K to remove the magnetic ordering before the next cooling phase of the M−H runs. The M-H sweeps used for constructing the phase diagrams are spaced at 1 K steps between 45 – 50 K for the $(Cu_{0.98}Co_{0.02})_2OSeO_3$ and $(Cu_{0.9}Co_{0.1})_2OSeO_3$ samples and 2 K steps between 46 – 50 K for the $(Cu_{0.95}Co_{0.05})_2OSeO_3$ sample. Meanwhile, between 50 and 60 K, the phase diagrams are spaced at 0.2 K for the $(Cu_{0.98}Co_{0.02})_2OSeO_3$ and $(Cu_{0.9}Co_{0.1})_2OSeO_3$ samples, and at 0.25 K for the $(Cu_{0.95}Co_{0.05})_2OSeO_3$ sample. The magnetic saturation measurements were performed at 2 K after ZFC cooling down from room temperature. Each sweep consists of an applied magnetic field ranging from 0 – 70 kOe, with 25 Oe step sizes between 0 – 4000 Oe, and then 500 Oe step sizes between 4000 – 70 kOe. The external magnet was then demagnetised in an oscillating fashion from 70 kOe to reduce the remnant magnetic field in the next M-H run.

### G. Small-Angle Neutron Scattering

The SANS experiments were conducted on QUOKKA, the monochromatic small-angle neutron scattering instrument at ACNS, ANSTO (Lucas Heights, Australia).[33,34] The experimental design, measurement parameters and data processing have been described elsewhere.[25] To ensure consistency across all measurements, a temperature stabilisation time of 5 minutes was incorporated to enhance the temperature accuracy and stability of each step. For magnetic field consistency, each sample was zero-field cooled to base temperature (4 K) before any magnetic field was applied perpendicular to the incident neutron beam, prior to taking measurements. To reset the magnetic system, the magnetic field was driven to zero (if one was applied) by oscillating the field before heating the sample to 120 K prior to cooling back to base temperature for the next set of measurements. Temperature sweep measurements had the applied magnetic field set either at 0 Oe, 200 Oe or 250 Oe before heating up in incremental temperature steps from 4 – 60 K. All the individual scans were measured for 5 minutes, except for the 0 Oe temperature sweep for $Cu_2OSeO_3$, which was done as 10 individual 60-second scans. A 5-minute time period was chosen to readily observe a clear peak intensity above the background.[34] Data were analysed by fitting Gaussian peaks to the resulting annular and radial peaks in the undoped sample, while for the Co-doped samples, Gaussian peaks were fitted to the annular peaks, and a Pseudo-Voigt fitting was used for the radial peaks, as it was observed that the peaks were narrower and provided a better fitting profile. These fittings yielded the maximum peak intensity in the annular data and the q-position of the peaks in the radial data, with a flat baseline.

The pXRD, SANS, and magnetic data for the undoped sample, for comparison, have been taken from our previous work with the permission of the authors.[25]

## III. Results and Discussion

Polycrystalline $(Cu_{1-x}Co_x)_2OSeO_3$ with different cobalt doping ($0 \leq x \leq 0.2$) were synthesised through solid-state synthesis. An elemental analysis was carried out using energy dispersive X-ray spectroscopy (EDS) and X-ray absorption spectroscopy (XAS) to confirm the presence of cobalt in the synthesised samples and that cobalt was in the +2 oxidation state, as shown in Table S1 and Fig. S1-S4. Additionally, neutron powder diffraction (NPD) showed that $Co^{2+}$ prefers to occupy the Cu2 site in the crystal structure of the $x = 0.1$ sample; however, in the $x = 0.2$ sample, it prefers the Cu1 site. The difference in the bound coherent neutron scattering lengths between cobalt and copper allows for modelling site occupancy from NPD data, unlike for synchrotron powder X-ray diffraction (pXRD) data.[35] The room temperature NPD patterns are shown in Fig. S6, with the Rietveld refinements presented in Fig. S7 and the Rietveld R-factors in Table S9.

The effects of cobalt substitution on the crystal structure can be studied using pXRD. Fig. 2 shows the high-resolution synchrotron pXRD patterns of the polycrystalline $(Cu_{1-x}Co_x)_2OSeO_3$ ($0 \leq x \leq 0.2$) samples at room temperature. The synthesised samples were confirmed to have high phase purity by refining the pXRD data against a $Cu_2OSeO_3$ model. The inset of Fig. 2 shows the shift in one of the main Bragg peaks to lower 2θ values due to the successful incorporation of $Co^{2+}$ ions into the crystal structure. This is represented by a systematic shift to the left with increasing $Co^{2+}$ doping as $Co^{2+}$ ions have a larger ionic radius of 0.67 Å when in a five-fold coordination geometry compared to $Cu^{2+}$ ions, 0.65 Å.[26]

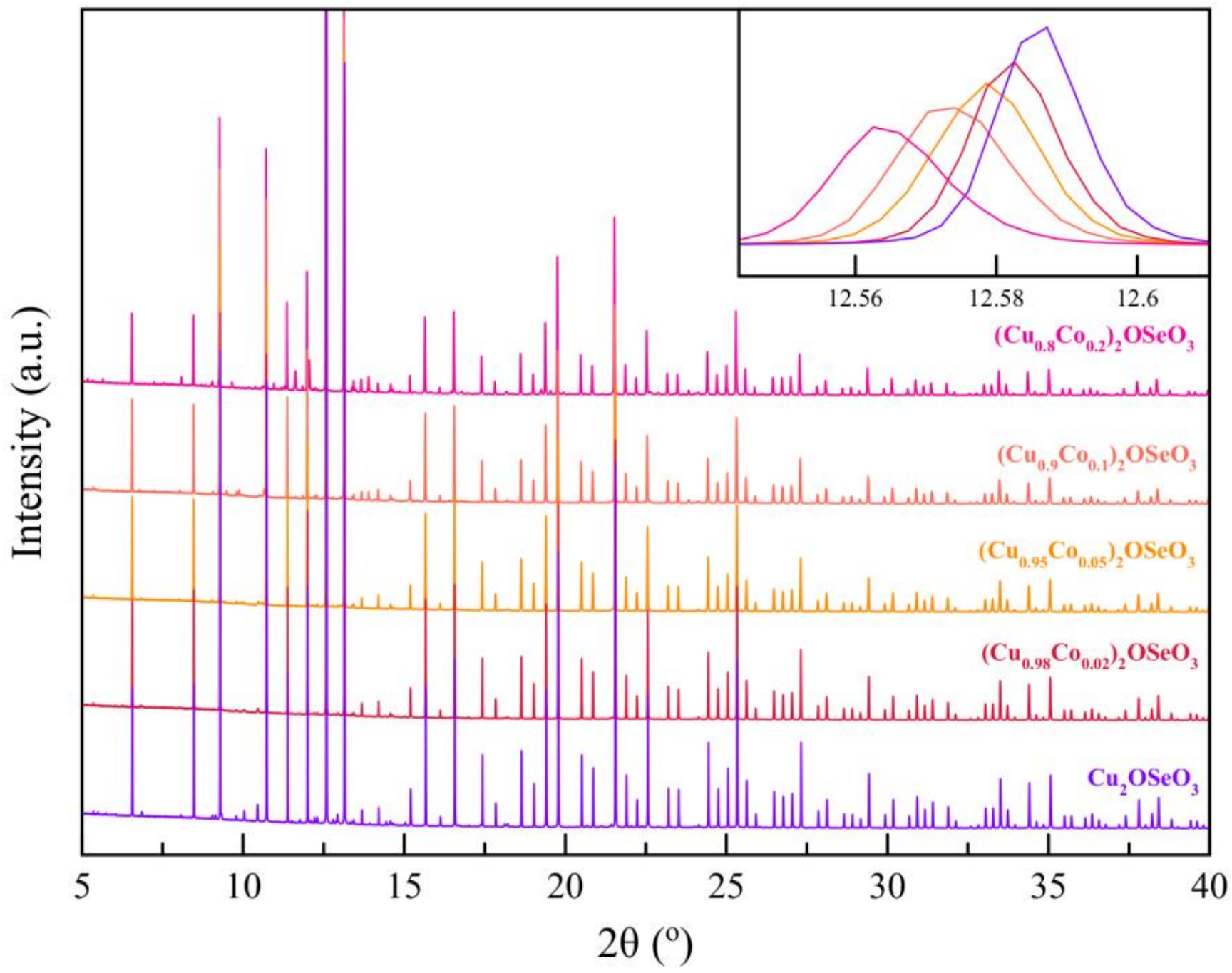

**FIG. 2.** Room temperature synchrotron pXRD patterns of the polycrystalline $(Cu_{1-x}Co_x)_2OSeO_3$ ($0 \leq x \leq 0.2$) samples with $x = 0$ at the bottom and $x = 0.2$ at the top, and shifted vertically for visual clarity. The data were collected using a wavelength of 0.59053(1) Å. The zoomed inset shows the systematic shift of one of the main Bragg peaks to lower 2θ as the nominal Co-doping level increases ($x$).

Rietveld refinements were carried out to gather structural information; the resulting unit cell parameters, bond lengths, bond angles and interspatial Cu-Cu distances are summarised in Tables S5-S6.[27–29] The Rietveld refinements are shown in Fig. S8. For the Co-doped samples, refinements were performed by fixing the $Co^{2+}$ occupancy at both $Cu^{2+}$ sites to match the nominal doping percentage, yielding better fitting values than when $Co^{2+}$ was not accounted for. The experimental data show a close fit to the calculated model for this system. The increase in lattice parameters with doping confirms the expansion of the unit cell with the successful incorporation of $Co^{2+}$ into the structure, showing a linear trend with increasing size, with greater incorporation of $Co^{2+}$ into the crystal structure as seen in Fig. S9.

As the spins of Cu1 and Cu2 align in an antiparallel arrangement, it is of specific interest to determine whether the $Co^{2+}$ ions have a preferential site since $Co^{2+}$ ions have a significantly stronger magnetic moment when they are in the high-spin configuration (3.87 μB compared to 1.73 μB for $Cu^{2+}$). If high-spin $Co^{2+}$ were to substitute into any of the three ferromagnetically aligned Cu2 sites, it would result in enhancing the overall magnetisation and ferromagnetic interaction. However, if it were to substitute for the Cu1 site, which is antiferromagnetically aligned with the Cu2 sites, it would decrease the overall magnetisation of the material, as shown in Fig. 3.

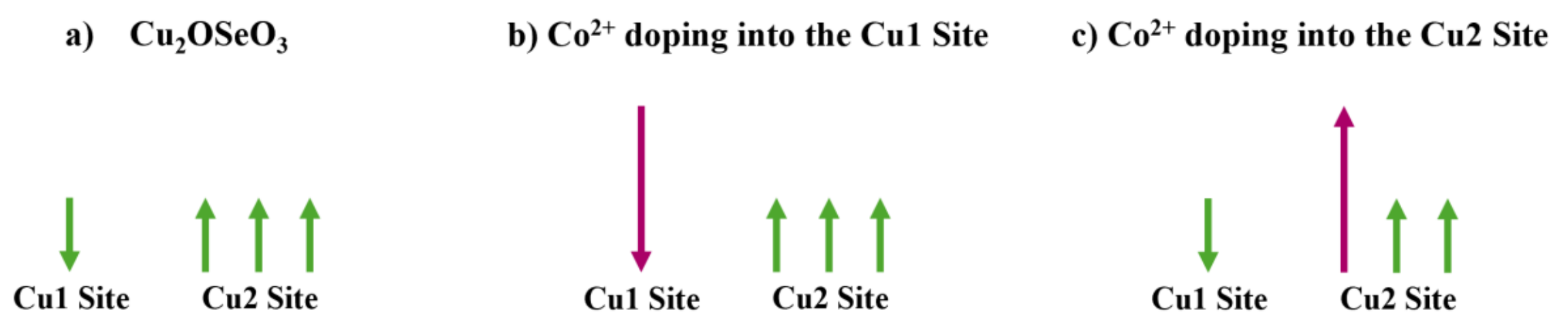

**FIG. 3.** A schematic representation of $Cu^{2+}$ spins in the Cu1 and Cu2 sites in a three-up one-down arrangement and the effects of $Co^{2+}$ doping into each site within $Cu_2OSeO_3$.

As cobalt and copper have very similar X-ray scattering lengths, the site occupancy of $Co^{2+}$ cannot be refined directly from X-ray data. Additionally, X-ray scattering has lower sensitivity for determining Cu-O bonds than neutron scattering. It can, however, be inferred indirectly from considering the changes in the respective interspatial Cu-Cu distances (Table S6) and $CuO_5$ polyhedra volume (Table S7) upon doping. Based on the pXRD and NPD refinements, there are no clear trends in the volumes of the Cu1 or Cu2 sites with increasing Co-doping. Meanwhile, the Cu1-Cu2 distances increase with increasing Co-doping while the Cu2-Cu2 distances do not, suggesting a preference for $Co^{2+}$ doping at the Cu1 site. This agrees with the Rietveld refinements of the NPD data for the $x = 0.2$ sample. This contrasts with $Zn^{2+}$ doping, which was shown to prefer to dope into the Cu2 site; meanwhile, $Ni^{2+}$ doping was suggested to have a random distribution in both copper sites.[18,19]

Upon the substitution of the $Co^{2+}$ atoms into the $Cu^{2+}$ sites, noticeable changes to the material's magnetic response can be observed from SQUID measurements. Fig. 4 shows the zero-field cooled (ZFC) temperature-dependent magnetisation data for the samples with $x \leq 0.1$, as the $x = 0.2$ sample contains magnetic impurities. The magnetic coils were quenched before cooling to eliminate the influence of an external remnant field. The ZFC temperature-dependent magnetisation measurement of the undoped sample can be seen in Fig. 4a. Using the inflection points from differentiating the measurement, two transitions are observed in the system. The higher temperature $T_C$ corresponds to the ordering transition from a paramagnetic (PM) to fluctuation disordered (FD) phase at $T_C = 58.97$ K. The second $T'_C$ occurs when the spins organise into a long-range helimagnet (HM) from the FD phase, ordering at $T'_C = 57.93$ K. This is similar to what has been previously observed for undoped $Cu_2OSeO_3$.[36,37] Upon substituting $Co^{2+}$ ions into the $Cu^{2+}$ site, the material remains a helimagnet, as evident by the characteristic magnetic transitions. However, both $T_C$ and $T'_C$ were observed to decrease in temperature with increasing $Co^{2+}$ content (Fig. 4b-d), with the values lowering to 58.81 and 57.48 K, 58.20 and 56.60 K, and 57.84 and 55.47 K for $(Cu_{1-x}Co_x)_2SeO_3$ with $x = 0.02$, 0.05, and 0.1, respectively. Such decreases in the $T'_C$ indicate a shortening in the effective SEI strength, J.[38] Another observation is that there is an increasing separation between the magnetic transitions in each of the Co-doped samples as $Co^{2+}$ content increases: $\Delta T = 1.04$, 1.33, 1.6 and 2.37 K for $x = 0$, 0.02, 0.05, and 0.1, respectively. The disproportionate lowering of the HM formation temperature alludes to the fact that the spin system is becoming more prone to thermally induced disorder as the $Co^{2+}$ doping increases, thus lowering the critical temperature at which long-range helimagnetic ordering is stable. When looking at the magnetic moment at the $T'_C$ transition, it can be observed that upon increasing $Co^{2+}$ inclusion, the specific moment also decreases, signifying a weakening in the strength of the transition along with more disordering of the helical domain. It should also be noted that for the $x$ = 0.1 sample, an additional feature can be seen immediately after the HM transition at lower temperatures in Fig. 4d, indicating the possibility of another magnetic transition being present.

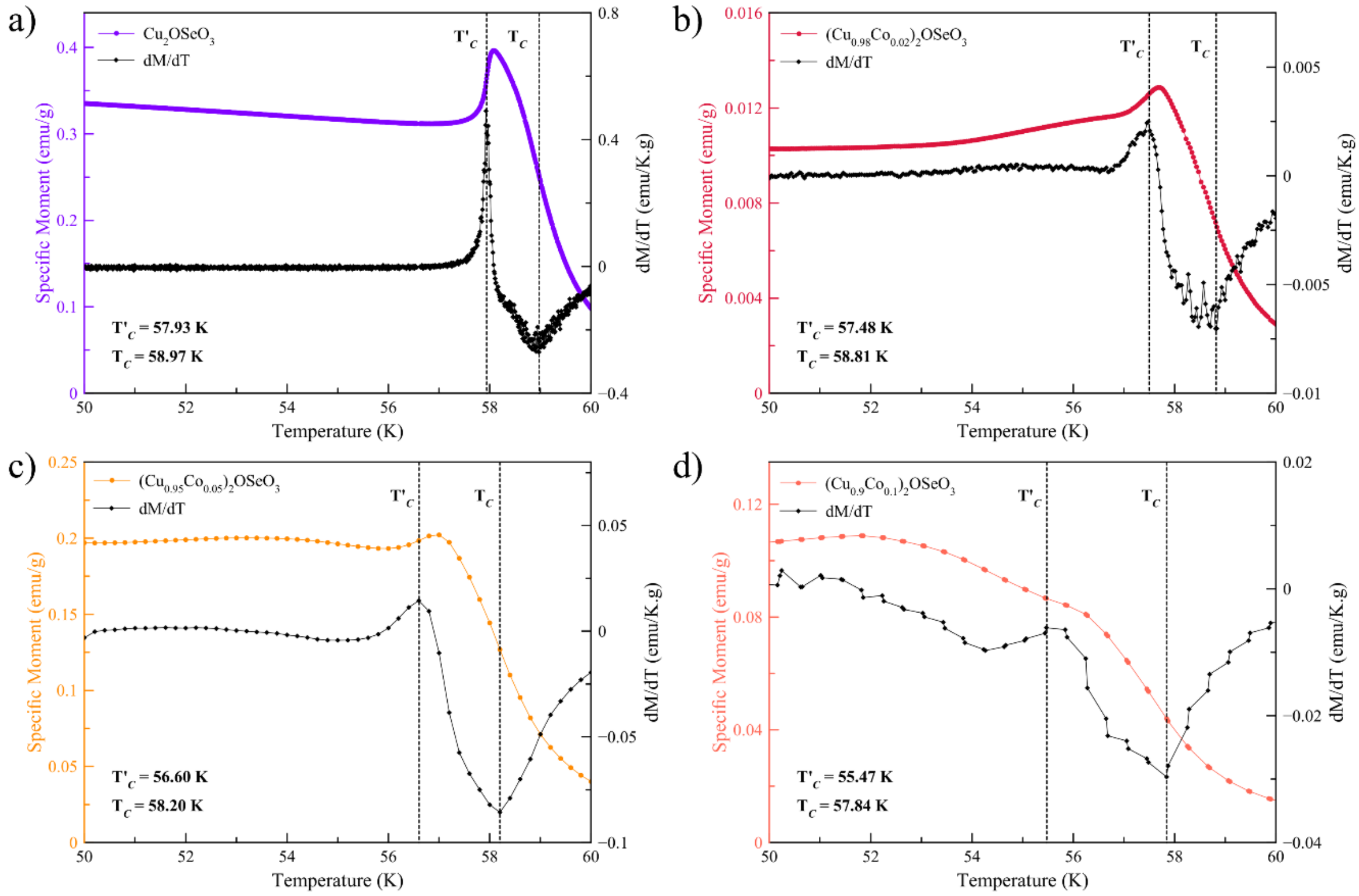

**FIG. 4.** Temperature-dependent magnetisation at zero magnetic field, along with their derivative, for the polycrystalline $(Cu_{1-x}Co_x)_2OSeO_3$ ($0 \leq x \leq 0.1$) samples measured between 2 – 80 K. The inflection points are shown representing the paramagnetic to fluctuation disordered transition ($T_C$) and the fluctuation disordered to helimagnetic transition ($T'_C$). a) $Cu_2OSeO_3$, b) $(Cu_{0.98}Co_{0.02})_2OSeO_3$, c) $(Cu_{0.95}Co_{0.05})_2OSeO_3$, and d) $(Cu_{0.9}Co_{0.1})_2OSeO_3$.

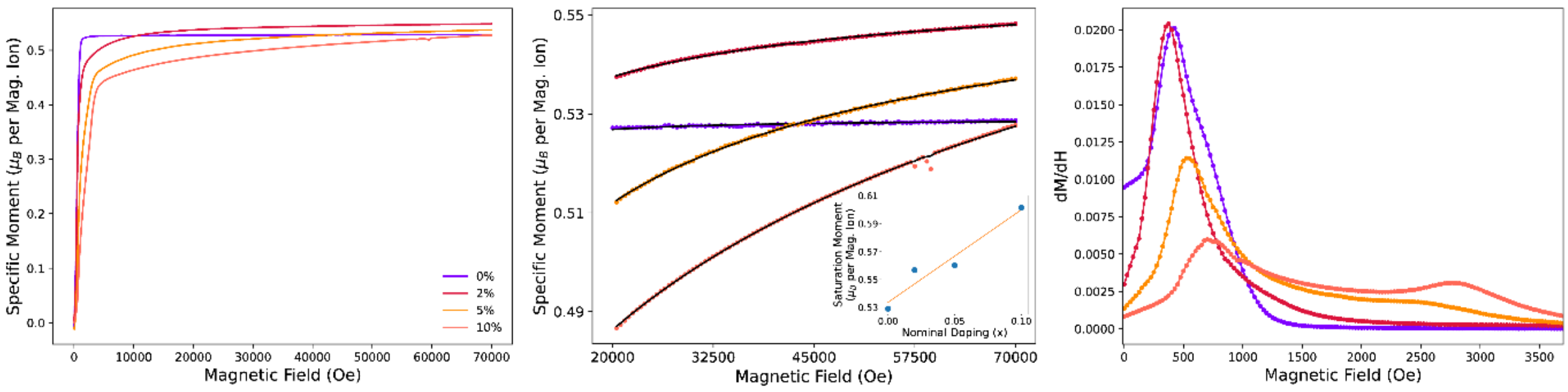

**FIG. 5.** Magnetisation in units of specific moment, μB per magnetic site, versus field in the polycrystalline $(Cu_{1-x}Co_x)_2OSeO_3$ ($0 \leq x \leq 0.1$) samples measured at 2 K. The saturation values have been corrected based on the phase purity of the sample as determined by synchrotron pXRD Rietveld refinements. The inset shows the extended Langevin model fittings for the magnetisation data for each sample. b) The first derivative of the specific moment $\mu_B/Cu^{2+}$ site against the magnetic field. The differential M-H data of the Co-doped samples were smoothed due to noisy data.

The influence of Co-doping on the magnetic spin interactions is further explored through low-temperature magnetisation measurements. Fig. 5a shows the field-dependent magnetisation measurements at 2 K with an applied field up to 70 kOe to observe where the magnetic saturation occurs. For the undoped sample, the magnetic saturation results in a value of 0.529 $\mu_B/Cu^{2+}$ per site; meanwhile, the $x \leq 0.1$ Co-doped samples have not yet reached saturation even at an applied field of 70 kOe. Instead, the high field section of the magnetisation measurements can be fitted with an extended Langevin model for anhysteretic ferromagnetism.[39] The form of the extended Langevin model is as follows:

$$M = M_s \left[ \coth\left(\frac{H + \beta M}{\alpha}\right) - \frac{\alpha}{H + \beta} \right]$$

Where the magnetisation, $M$, is a function of the saturation magnetisation, $M_s$, magnetic field, $H$, β is the cluster interaction parameter, and $\alpha = \frac{k_B T}{\mu_0 m}$, where $k_B$ is the Boltzmann constant, $\mu_0$is the vacuum permeability, $T$ is the temperature, and $m$ is the cluster magnetisation. This model accounts for the slow rate of increase in comparison with the Langevin model by describing a system of magnetic clusters or domains, with smaller clusters or lower inter-cluster interactions resulting in a higher saturation field. These fitting parameters are depicted in Fig. S10, where in a) the inter-cluster interaction parameter rapidly increases with doping, while in b) the cluster magnetisation decreases from 358(96) $\mu_B$ for $x = 0$ to 2(9) $\mu_B$ for $x = 0.1$, implying a transition from small magnetic clusters to atomic spin like units on the scale of the individual strongly coupled polyhedra. This strongly suggests that a decrease in the inter-tetrahedral SEI coupling is driving the decrease in cluster size with doping. The result is a higher saturation field, which can be found by iteratively solving for self-consistency of the extended Langevin model under the saturation condition $M(H_{sat}) = 0.975M_{sat}$. The saturation fields appear to increase dramatically with doping, as shown in Fig. S10c, with the saturation field values of 3.31, 35.83, 133.78, and 682.42 kOe for $x = 0$, 0.02, 0.05, and 0.1, respectively.

Fig. 5b) shows the high field portion of the magnetisation curves, fitted by the extended Langevin model. The inset depicts the resulting saturation magnetisation, with saturation values of 0.529, 0.557, 0.560, and 0.601 $\mu_B$ per magnetic ion for $x = 0$, 0.02, 0.05, and 0.1, respectively. This growth in moment per ion indicates the $Co^{2+}$ ions are preferentially doped onto the Cu2 site, at least at a higher ratio than into the Cu1 site. This agrees with the Rietveld refinements of the NPD data for the $x = 0.1$ sample. Assuming the $Co^{2+}$ ions are doping into sites with the same ratio irrespective of cobalt doping content, the linear fit corresponds to a $Co^{2+}$ moment of 1.73 $\mu_B$ per magnetic ion for doping solely into the Cu2 site, or 2.40 $\mu_B$ per magnetic ion for the 3:1 ratio of the Cu2 to Cu1 sites. Although there is insufficient statistical accuracy in the pXRD data to determine the precise position of the $Co^{2+}$ spin in these systems, the exclusively Cu2 site distribution corresponds exactly with the low spin configuration moment of $Co^{2+}$, while the 3:1 ratio site distribution would imply either a mixture of spin configurations or a high spin moment significantly modified by the nature of the local environment. Combining magnetisation and NPD measurements suggests that at lower Co-doping concentrations ($x \leq 0.1$), cobalt prefers the Cu2 site, whereas at higher Co-doping ($x = 0.2$), it prefers the Cu1 site.

More information can be extracted from the M-H curves by considering the differential susceptibility, d*M*/d*H*, as shown in Fig. 5c. A notable change is the progressive increase of the critical fields, $H_{C1}$ and $H_{C2}$, which are the fields required for the transition from helical to the conical phase and from the conical to the field-polarised phase, respectively.[40] In Fig. 5c, a large peak is located at 400 Oe for the undoped sample, with a small shoulder corresponding to the metamagnetic transition found at lower temperatures.[40] At $x = 0.02$, this shoulder disappears along with a shift in the main peak to lower fields and intensity, pointing to a possible change in the spin flip behaviour driving the metamagnetism. With further increases in Co-doping, both the $x = 0.05$ and $x = 0.1$ samples clearly show a shift in the first peak, which broadens, shifts to higher fields, and decreases in intensity as the doping increases. This implies that a greater magnetic field is required for the SEI to overcome the canting due to DMI and transition to the field-polarised phase. Interestingly, for the $x = 0.05$ and $x = 0.1$ samples, there is the emergence of a new shoulder at approximately 2500 and 2775 Oe in the derivative, respectively. This signifies a change in the nature of the conical to field-polarised state, potentially including a new phase that is not present in the other samples. The skyrmion temperature stability range can also be investigated through mapping the inflection points in the differential susceptibility by magnetometry, as shown in Fig. 6. From this, the critical fields for the helical to conical, conical to skyrmion, and skyrmion to conical are found and labelled as $H_{C1}$, $H_{S1}$, and $H_{S2}$, respectively. The critical fields are shown on the stack plots in Fig. 6(a1-c1). The critical fields are overlaid on the magnetic phase diagram to highlight where the transitions are occurring around the skyrmion pocket shown in Fig. 6(a2-c2). It can be clearly observed that Co-doping significantly affects the temperature and field conditions required for skyrmions to be stabilised. With increasing Co-doping, skyrmions form at lower temperatures and higher fields, which agrees with the M-T data. Based on the magnetometry data, the skyrmions are present at 48 – 57.4, 50 – 56.75, and 49 – 56 K for the $x = 0.02$, 0.05, and 0.1, respectively. This shows that the skyrmion pocket has expanded and stabilised over a wider temperature range compared to 56.8 – 58.2 K in the undoped sample.[25] The skyrmion formation temperature does not follow a linear shift with increasing Co-doping; however, the skyrmion dissolution temperature shifts to lower temperatures with increasing Co-doping, following a similar trend to the M-T data. Despite this, a small amount of Co-doping shows a significant expansion of the skyrmion pocket temperature range.

The skyrmion field stability range also shifts to higher fields as the pocket widens with increasing Co-doping. Taking the widest field range between the critical fields $H_{S1}$ and $H_{S2}$, in the $x = 0.02$ sample, at 53 K, the skyrmions are stable between 220 – 380 Oe. With increasing Co-doping, the skyrmion field stability increases with $H_{S1}$ shifting to lower fields and $H_{S2}$ shifting to higher fields, with the largest field range being situated at 55 K for both the $x = 0.05$ and 0.1 samples, with a respective field range of 180 – 390 and 130 – 400 Oe. In comparison, the widest field range for the undoped sample was found to be between 125 – 250 Oe at 56 K for $x = 0.1$. This shows that, with increasing Co-doping, the skyrmions are stabilised over a wider field range, extending to higher fields. This trend of increasing stability of skyrmions to lower fields is unique amongst Co-doped samples, as other dopants have increased the skyrmion pockets to higher fields but not to the lower fields as well.[25]

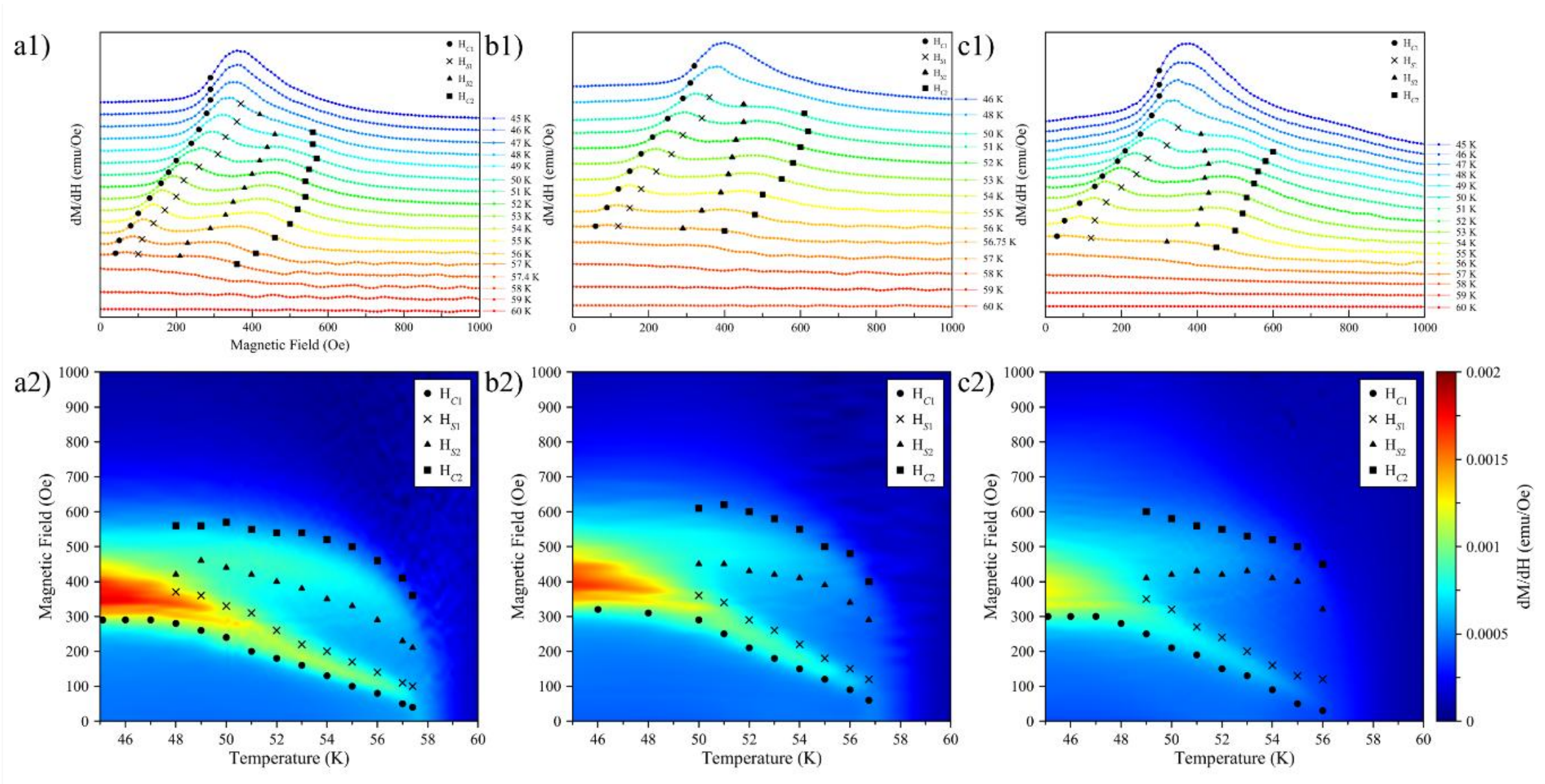

**FIG. 6.** The magnetic phase diagrams for the polycrystalline $(Cu_{1-x}Co_x)_2OSeO_3$ $(0.02 \leq x \leq 0.1)$ samples between 45 – 60 K and 0 – 1000 Oe field. a) $(Cu_{0.98}Co_{0.02})_2OSeO_3$, b) $(Cu_{0.95}Co_{0.05})_2OSeO_3$,

and c) $(Cu_{0.9}Co_{0.1})_2OSeO_3$. The inflection points of the differential susceptibility are plotted as a function of temperature, shown in the first row of the figure (a1, b1, c1), corresponding to the critical field points. These points correspond to the transitions, helical to conical, conical to skyrmion, skyrmion to conical, and conical to ferrimagnetic, labelled as $H_{C1}$, $H_{S1}$, $H_{S2}$, and $H_{C2}$, respectively. These critical points are superimposed onto the magnetic phase diagrams of the corresponding samples shown in the second row of the figure (a2, b2, c2).

| **Nominal Composition (*x*)** | **Critical Temperatures (K)** | | **Saturation** | | **Critical Fields (Oe)** | | **Skyrmion Pocket (K)** | |
|---|---|---|---|---|---|---|---|---|
| | $T_C$ | $T'_C$ | Saturation Fields (kOe) | Saturation Magnetisation ($\mu_B$ per magnetic ion) | $H_{S1}$ | $H_{S2}$ | Magnetisation Data | SANS Data |
| 0 | 58.97 | 57.93 | 3.31 | 0.529 | 125 | 250 | 56.8 – 58.2 | 55 – 57.5 |
| 0.02 | 58.81 | 57.48 | 35.83 | 0.557 | 220 | 280 | 48 – 57.4 | 45 – 56 |
| 0.05 | 58.20 | 56.60 | 133.78 | 0.560 | 180 | 390 | 50 – 56.75 | 4 – 60 |
| 0.1 | 57.84 | 55.47 | 682.42 | 0.601 | 130 | 400 | 49 – 56 | - |

**TABLE I.** Summary of the magnetic and skyrmion-related parameters for the $(Cu_{1-x}Co_x)_2OSeO_3$ ($0 \leq x \leq 0.1$) samples. The table lists the critical temperatures, $T_C$ and $T'_C$, saturation fields, and saturation magnetisation. Also reported are the critical fields, $H_{S1}$ and $H_{S2}$, and the skyrmion pocket temperature ranges determined from the magnetisation and SANS measurements.

To confirm that the magnetic transitions observed from the SQUID data after the inclusion of $Co^{2+}$ into the crystal structure are indeed helimagnetic, small-angle neutron scattering (SANS) measurements were carried out. As shown in our previous work, different helimagnetic phases in a polycrystalline sample can be distinguished by their unique magnetic SANS scattering patterns.[25] In the absence of an applied magnetic field, the material exhibits a helical ordering when cooled below the critical temperature, $T_C$. For all samples, a zero-field temperature sweep was measured to investigate how incorporating cobalt affects the helimagnetic nature of the samples. Videos compiled from the detector images of this zero-field temperature measurement can be seen in Movie S1. The patterns found in each frame were analysed individually to quantitatively assess the influence of $Co^{2+}$ on the helimagnetic structures within the material. As shown in Fig. 7a, all the polycrystalline samples exhibit helimagnetic ordering at 0 Oe, with a decrease in the helical ordering intensity upon Co-doping. The intensity of the helical phase shows a temperature dependency characteristic of a second-order phase transition, with the intensity decreasing towards zero as the samples were heated towards the $T_C$.[41] The helimagnetic ordering has also been observed to remain for other dopants in the magnetic site, such as nickel and zinc.[23] Unlike what was observed with non-magnetic doping with tellurium[25], a decrease in radial peak intensity corresponding to the helical phase with increasing amounts of Co-doping can be seen, along with a shift to lower temperatures for the helical to FD phase transition. The helical phase being present at a higher temperature for the $x = 0.02$ sample at 57 K compared to 56 K for the undoped sample is caused by a small change in an experimental regime, as the undoped sample was only measured in 2 K temperature steps at higher temperatures, while the Co-doped samples were measured in 1 K temperature steps. The helical phase likely persists to 57 K in the undoped sample, but it was not observed at 58 K, when it is in the FD regime. The downwards shift of 1 K from 57 K in the $x = 0.02$ sample to 56 K in the $x = 0.05$ sample agrees with the magnetisation data. The intensity of the scattering patterns provides insight into the number of magnetic domains present in the material. As seen with increasing Co-doping, the scattering intensity decreases, indicating that fewer magnetic domains form in the Co-doped samples. In this case, with increasing Co-doping, fewer helical domains are formed throughout the zero-field temperature sweep.

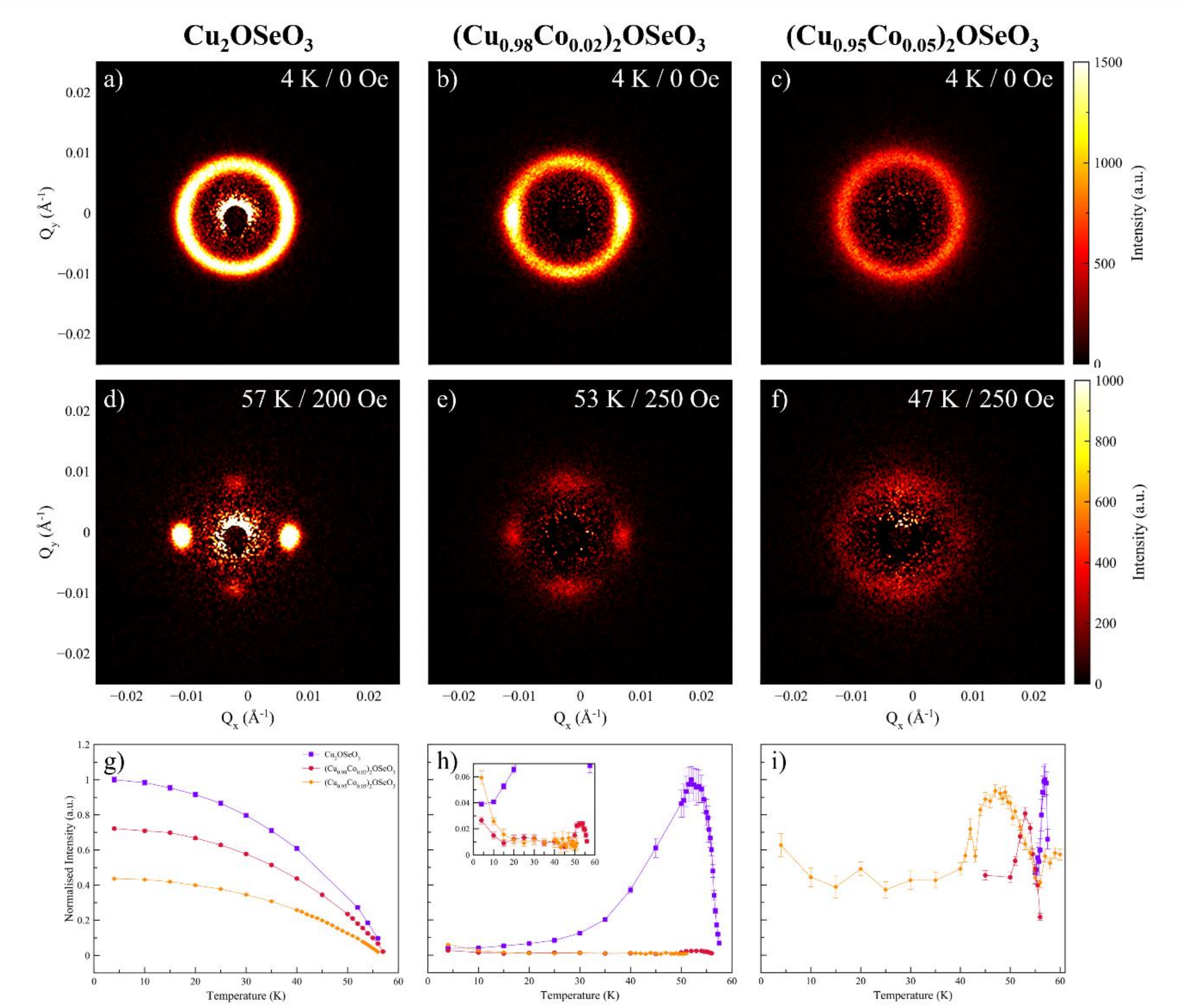

**FIG. 7.** The SANS detector images and temperature-dependent peak intensities for the polycrystalline $(Cu_{1-x}Co_x)_2OSeO_3$ ($0 \leq x \leq 0.0.5$) samples. (a-c) Zero-field detector images at 4 K showing the helical phase for $Cu_2OSeO_3$, $(Cu_{0.98}Co_{0.02})_2OSeO_3$, and $(Cu_{0.95}Co_{0.05})_2OSeO_3$. b) The conical phase can be observed due to a remanent magnetic field. (d-f) Field-dependent images showing the conical and skyrmion phases at 200 Oe ($Cu_2OSeO_3$) and 250 Oe (Co-doped samples) at 57, 53, and 47 K, respectively. (g) Radial peak intensity at zero field (helical phase). (h) Annular peak intensity at 0° and 180° (conical phase); the inset shows the enlarged conical intensities for the Co-doped samples. (i) Annular peak intensity at 90° and 270° (skyrmion phase).

Upon applying an external magnetic field of up to 200 Oe for the undoped sample and 250 Oe for both Co-doped samples, the conical and skyrmion phases appear. Their evolution is shown in Supplementary Movie S1. The conical feature is depicted as two intense $q_y = 0$ Bragg peaks at the same |q| located at 0º and 180º, and their analysis is shown in Fig. 7b, with the inset showing the zoomed-in conical intensity of the two Co-doped samples. It is clearly observed that Co-doping results in a large suppression of the conical phase from transitioning out of the helical phase, as seen in the undoped sample. Looking at the inset, the $x = 0.02$ sample still shows a conical phase peak maximum at 54 K, which is 2 K higher, while the x = 0.05 sample shows none. For the $x = 0.05$ sample, a suppression of the conical phase at higher temperatures is apparent. There is a maximum conical intensity at 4 K, followed by a rapid decrease to 20 K, where it remains at approximately 1% of the undoped conical intensity until 51 K, after which it disappears. The $x = 0.02$ sample also has this feature with the conical phase having the highest intensity at 4 K before decreasing to 15 K which remains relatively constant until 45 K before increasing in intensity again and then disappearing at 56 K. Despite no apparent trend in the conical phase peak maxima shifting to lower temperatures there is a shift to lower temperatures when the conical phase disappears. The conical phase disappears at 57.5 K in the undoped sample, 56 K in the $x$ =0.02 sample, and 51 K in the $x = 0.05$ Co-doped sample, which is a significant shift of 6.5 K.

Lastly, in Fig. 7c, the skyrmion phase shows as two $q_x = 0$ peaks at the same |q| located at 90º and 270º. As is well known for $Cu_2OSeO_3$, the skyrmion phase evolves from the conical phase at higher temperatures, with both phases coexisting before disappearing and transitioning into the FD regime.[25] The coexistence of the conical and skyrmion phases also remains with Co-doping; however, a significantly greater volume fraction of the skyrmion phase is formed compared to the conical phase. The maximum skyrmion volume fraction is 52(2)% for the $x = 0.02$ sample at 53 K and 100% between 52 – 60 K for the $x = 0.05$ sample. Compared to the undoped sample, there is only a skyrmion volume fraction of 24(2)%, as shown in Fig. S11. Compared to magnetisation data, the $x$ = 0.02 sample hosts skyrmions across a wider temperature range; however, for the $x = 0.05$ sample, the skyrmions are observed over a significantly larger temperature range. Despite the effects of the

metastable skyrmions, the peak feature corresponding to the skyrmion intensity as a function of temperature is apparent in the $x = 0.05$ sample at 40 K in Fig. 7c. This indicates that with increasing Co-doping, the formation of skyrmions occurs at lower temperatures and over a larger temperature range. This highlights the fact that Co-doping has an effect on the magnetic exchange interactions, so that less thermal energy is required to nucleate the skyrmion phase. Another interesting observation is that, for the $x = 0.05$ sample only, the skyrmions are stable at 9 K higher than when the conical disappears in the sample, whereas in both the undoped and $x = 0.02$ samples the skyrmion disappears at the same temperature as the conical. The shift of the skyrmion stability to lower temperatures is characteristic of an expansion of the crystal structure, as seen with non-magnetic Te-doping.[25]

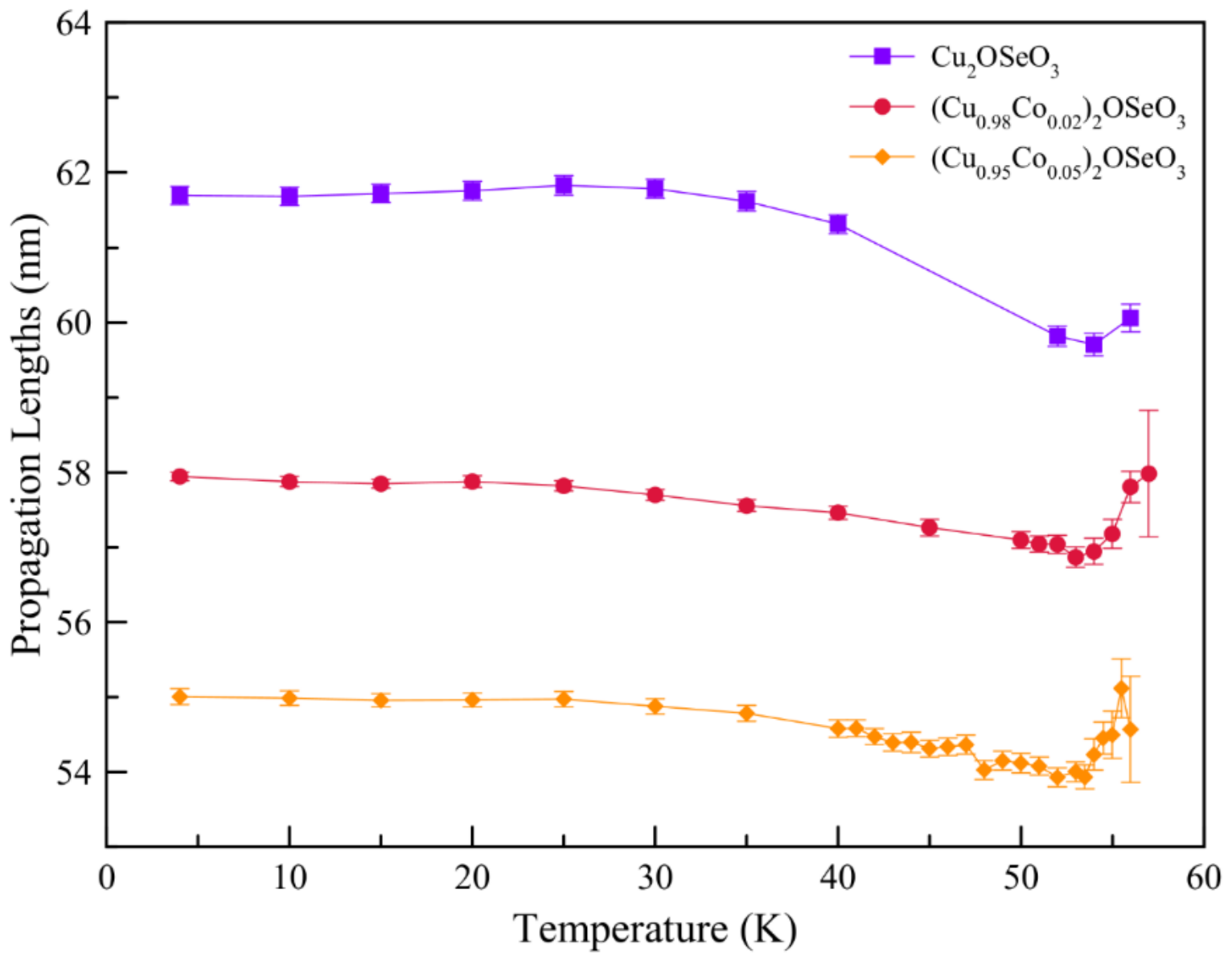

**FIG. 8.** The helical ordering propagation lengths of the polycrystalline $(Cu_{1-x}Co_x)_2OSeO_3$ $(0 \leq x \leq 0.05)$ samples determined from the integrated SANS radial peak q-positioning, with zero applied magnetic field as a function of temperature.

As shown in the literature and in our previous work in SANS,[23,25,40–43] the radial peak q-position is observed at 0.01020(2) $Å^{-1}$ for the undoped sample at 4 K. Upon Co-doping, a significant shift in the q-position of the radial peak to greater q-values with increasing amounts of Co-doping becomes evident. Upon 2% doping ($x = 0.02$), radial peak position increases to 0.01090(1) $Å^{-1}$, followed by another increase to 0.01140(2) $Å^{-1}$ upon 5% doping ($x = 0.05$). As the q-position of the peak is inversely related to the magnetic ordering propagation length, there is a decrease in the propagation length of the helical ordering. As shown in Fig. 8, the helical propagation length, λ, decreases from 61.7(1), to 57.9(1), to 55.0(1) nm for $x = 0$, 0.02, and 0.05, respectively, at 4 K. Interestingly, it would be expected that upon lattice expansion from Co-doping the propagation lengths would increase similarly to what was observed in Te-doped sample but this is not observed here.[25] Instead, it appears that the lattice expansion effect is outcompeted by a change in the magnetic interaction strength. The helical propagation length relates to the SEI and DMI strengths through $\lambda \propto J/D$, suggesting that the SEI strength has decreased relative to the strength of the DMI with Co-doping, consistent with our magnetometry analysis.[14] This suggests that upon Co-doping, the full helical rotation happens across a smaller distance and shortens the helical propagation length. The propagation lengths also display a maximum at approximately 30 K. This temperature corresponds to the observation of a similar peak in the temperature-dependent magnetisation, which is attributed to a strain-induced change in inter-polyhedral SEI strengths.[40] In temperature-dependent magnetisation measurements, this results in the crossover of one inter-polyhedral SEI coupling from ferromagnetic to antiferromagnetic, thereby reducing the total magnetisation. In the case of the propagation length, the effective total exchange strength J could be reduced by this mechanism, thereby reducing the propagation length.

## IV. Conclusion

We have synthesised and characterised Co-doped polycrystalline $(Cu_{1-x}Co_x)_2OSeO_3$ ($0 \leq x \leq 0.2$) samples. This work provides evidence that $Co^{2+}$ ions can be successfully doped into the crystal structure, as confirmed by a combination of XAS, EDS, pXRD, and NPD refinement analysis. Upon doping, the unit cell volume increases, and cobalt shows a preference for occupying the Cu2 site for the $x \leq 0.1$ sample before changing to the Cu1 site in the $x = 0.2$ sample. The preferential doping led to a decrease in the strength of the SEI relative to the cobalt content, as seen from the magnetisation measurements. This is evident from the decreases in both the critical temperatures and a broadening in their separation between $T_C$ and $T'_C$ with increasing Co-doping. As a result of the drop in SEI, we also observed a decrease in ferrimagnetic cluster size. This led to an increase in the saturation field with doping. At 2 K, the $x = 0.1$ sample did not reach saturation even at 70 kOe. Furthermore, Co-doping increased the critical fields for various magnetic transitions to higher fields, along with the emergence of a new magnetic transition in the $x = 0.05$ and $x = 0.1$ samples at 2500 and 2775 Oe, respectively. SANS confirmed that Co-doping does not break the helimagnetic ordering in the system, despite changes in the magnetic exchange interaction. However, it reduces the number of magnetic domains present, as shown by the lower scattering intensity of the phases. Skyrmions were also observed at lower temperatures and higher fields with increasing Co-doping. The decrease in the required field for skyrmion formation led to the system being more susceptible to the formation of metastable skyrmions. This was observed in the $x = 0.05$ sample, despite applying a similar "zero field" as in other samples, metastable skyrmions persisted from 4-60 K. This work presents an interesting avenue to control the magnetic exchange interaction through preferential insertion of magnetic ions. It is clear that, despite the persistent presence of skyrmion lattices and other helimagnetic orderings, their formation and dynamics can be tuned by extrinsic magnetic doping. It will be of interest for subsequent in-depth experiments to understand better the effects of Co-doping on the stability of the various magnetic orderings, and to probe the novel magnetic transition observed at 2 K, especially at higher magnetic fields.

## Acknowledgements

We thank Dr. L. Tan for collecting the room temperature synchrotron powder X-ray diffraction data at the Australian Synchrotron, a part of the Australian Nuclear Science and Technology Organisation (ANSTO). We also thank Dr. J. Wykes for his assistance as our beamline scientist in during our MEX1 synchrotron beamtime. We also would like to thank the ANSTO ACNS sample environment team: G. Davidson, T. d'Adam and C. Baldwin, for their continuous support over our SANS experiment (Proposal No. P15924). This research was funded by the Royal Society Te Apārangi Marsden Fund (20-UOA-225), AINSE Ltd. Postgraduate Research Award (PGRA) (ALNSTU13239) and the University of Auckland Doctoral Scholarship for M. Vás. We also acknowledge the support through the Australian Research Council (ARC) through the funding of the Discovery Grant DP170100415 and the funding of the Linkage Infrastructure, Equipment and Facilities Grant LE180100109. This research was undertaken in part on the Powder Diffraction beamline at the Australian Synchrotron, part of ANSTO (Proposal No. PDR21589), on the Medium Energy X-ray Absorption Spectroscopy beamline at the Australian Synchrotron (Proposal No. M21926), as well as on the QUOKKA instrument (Proposal No. P15924) and ECHIDNA beamline (Proposal No. MI17267) at the Australian Centre for Neutron Scattering (ACNS) at ANSTO.

## Author Contributions

MV, SY, CU and TS conceived the project idea. MV synthesised all the samples and performed structural, elemental and neutron scattering characterisation. MV and HMC performed NPD measurements. MV, SY and TS performed XAS measurements. AJF, SY and CU performed magnetic measurements. MV, EPG and SY performed the SANS measurements. TS performed the analysis of the XAS data. MV, AJF and SY performed the analysis of the magnetometry data. MV, AJF and SY wrote the manuscript. All authors contributed to the discussion of the results and the improvement of the manuscript. SY and TS supervised the project.

## Data Availability

The data that support the findings of this article are not publicly available. The data are available from the authors upon reasonable request.

**Supplementary Information: The Effects of Cobalt Doping on the Skyrmion Hosting Material, $Cu_2OSeO_3$**

| Nominal Composition ($x$) | Atomic Percentage (%) | | | | Co-Doping (%) |
|---|---|---|---|---|---|
| | O $K$ | Cu $K$ | Co $K$ | Se $L$ | |
| **0.02** | 58.36 | 26.74 | 0.63 | 14.27 | 2.30 |
| **0.05** | 60.16 | 24.08 | 1.52 | 14.24 | 5.94 |
| **0.1** | 57.23 | 24.44 | 2.98 | 15.37 | 10.46 |
| **0.2** | 54.00 | 22.56 | 7.80 | 15.62 | 25.69 |

**TABLE S1.** EDS data collected for $(Cu_{1-x}Co_x)_2OSeO_3$ ($0.02 \leq x \leq 0.2$) samples. Composition quantified through the oxygen, copper, cobalt $K$-edges and selenium $L$-edge. The determined overall doping of cobalt occupation into the copper sites is shown in terms of percentage.

The EDS spectra and their corresponding SEM images are the EDS spectra can be found in Fig. S1-S4. EDS was able to detect 2.30% cobalt in the $x = 0.02$ sample and 5.94% cobalt in the $x = 0.05$ sample. The possible reason why both Co-doped samples have a slightly higher percentage of cobalt than the nominal amount could be due to the decomposition of the material near the end of the synthesis, resulting in a higher ratio of $Co^{2+}$ to $Cu^{2+}$ in the structure. It has been observed that if the reaction vessel is held long enough at the synthesis temperature, the material can start to decompose into a 2:1 molar ratio of CuO and $SeO_2$.[44] Meanwhile, the EDS for the $x = 0.1$ sample was able to detect 10.46% cobalt in the sample. Since this sample has Co-based impurities estimated to be between 1 – 2%, the actual Co-doping could be somewhere between 8 – 9%. Lastly, for the $x = 0.2$ sample, EDS detected 25.69% cobalt in the sample. According to the phase percentages from the synchrotron pXRD data, there was also approximately 4% of $CoSeO_3$ and 1% of $Co_3O_4$ impurity phases in the sample; however, the elemental ratios at% agree that the combination of cobalt and copper is 2:1, with selenium indicating that the measured crystallites were the correct phase rather than these impurities.

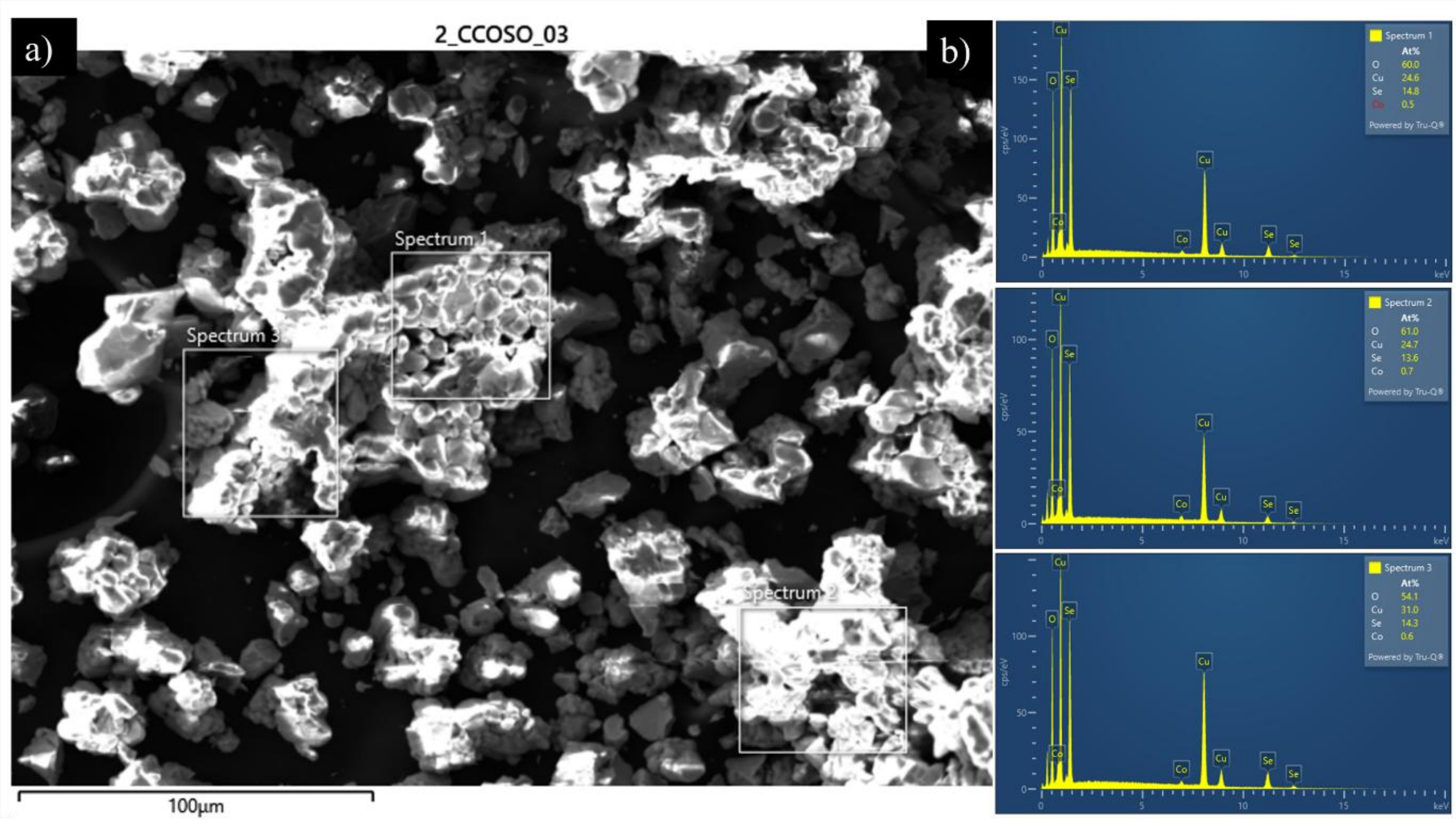

**FIG. S1.** a) A backscattered SEM image of the polycrystalline $(Cu_{0.98}Co_{0.02})_2OSeO_3$ sample with three white boxes identifying the regions that were scanned with EDS. The image was measured under a WD of 9.3 mm and a magnification of 300x. b) The three corresponding EDS spectra.

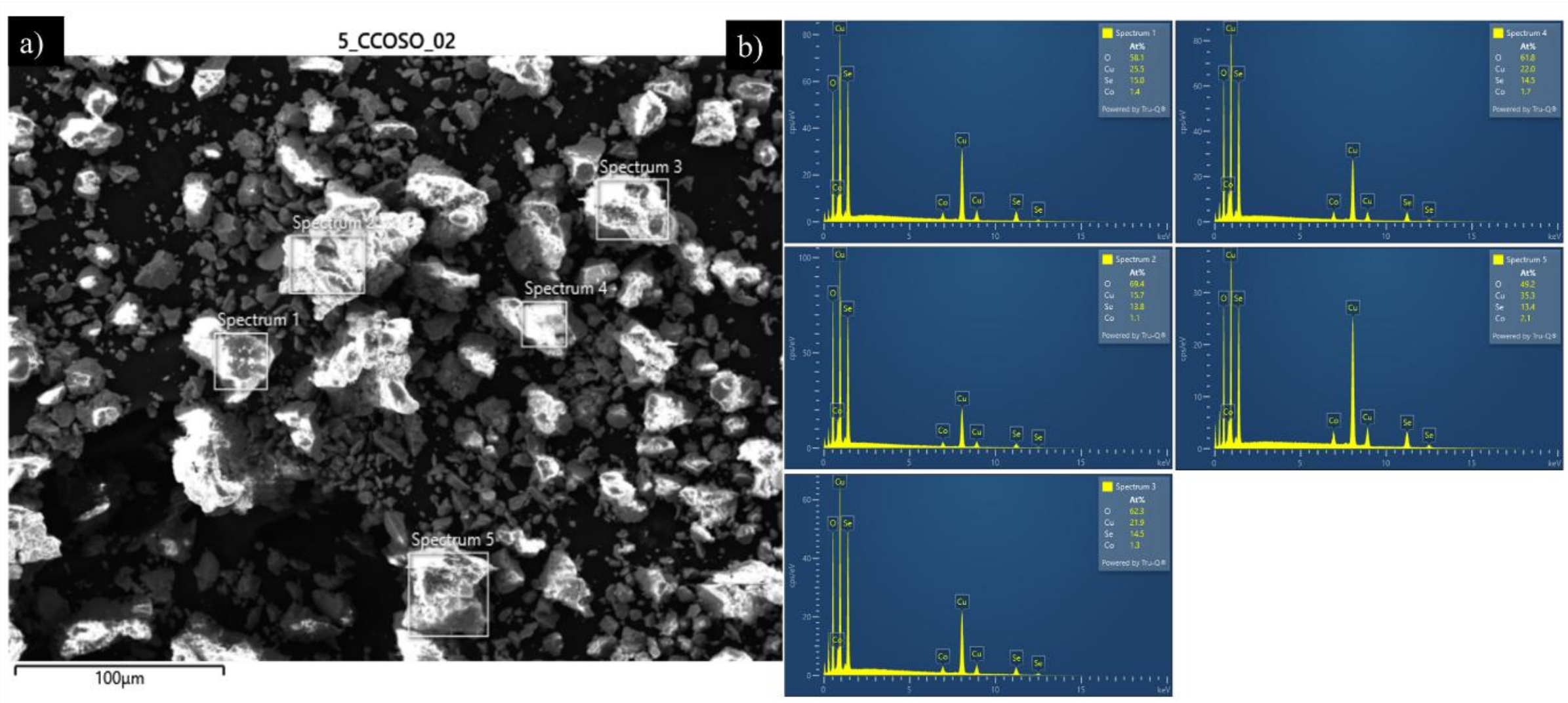

**FIG. S2.** a) A backscattered SEM image of the polycrystalline $(Cu_{0.95}Co_{0.05})_2OSeO_3$ sample with five white boxes identifying the regions that were scanned with EDS. The image was measured under a WD of 9.2 mm and a magnification of 311x. b) The five corresponding EDS spectra.

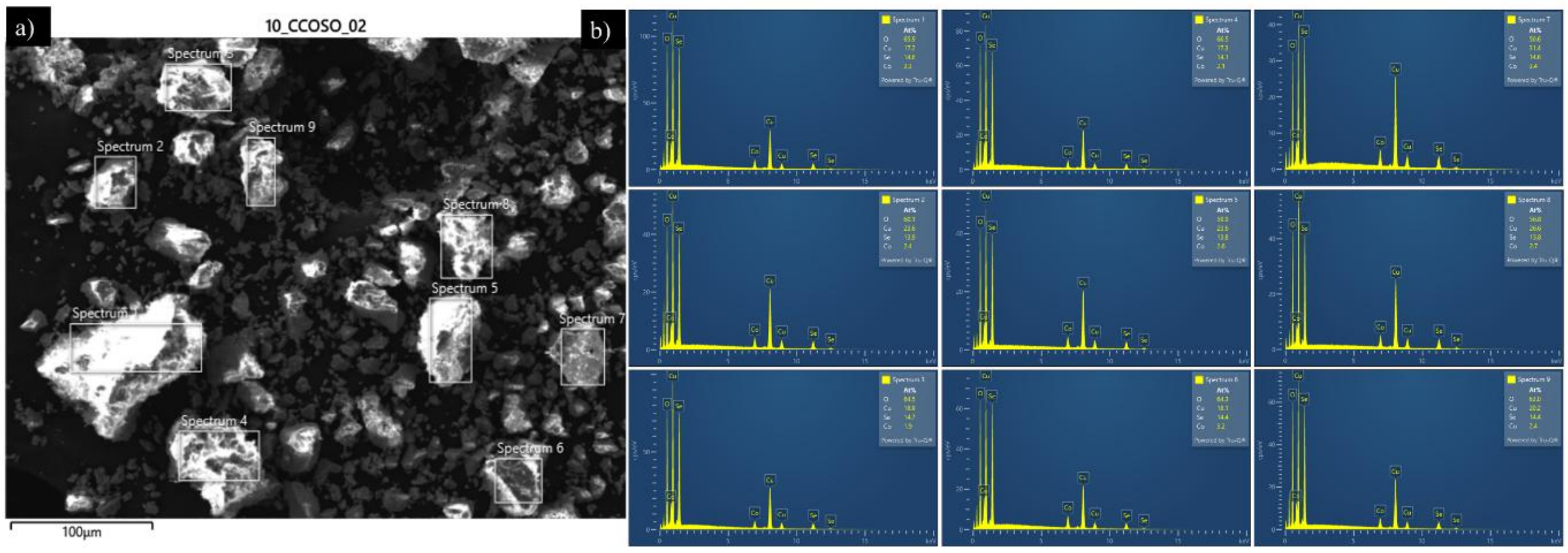

**FIG. S3.** a) A backscattered SEM image of the polycrystalline $(Cu_{0.9}Co_{0.1})_2OSeO_3$ sample with nine white boxes identifying the regions that were scanned with EDS. The image was measured under a WD of 9.3 mm and a magnification of 264x. b) The nine corresponding EDS spectra.

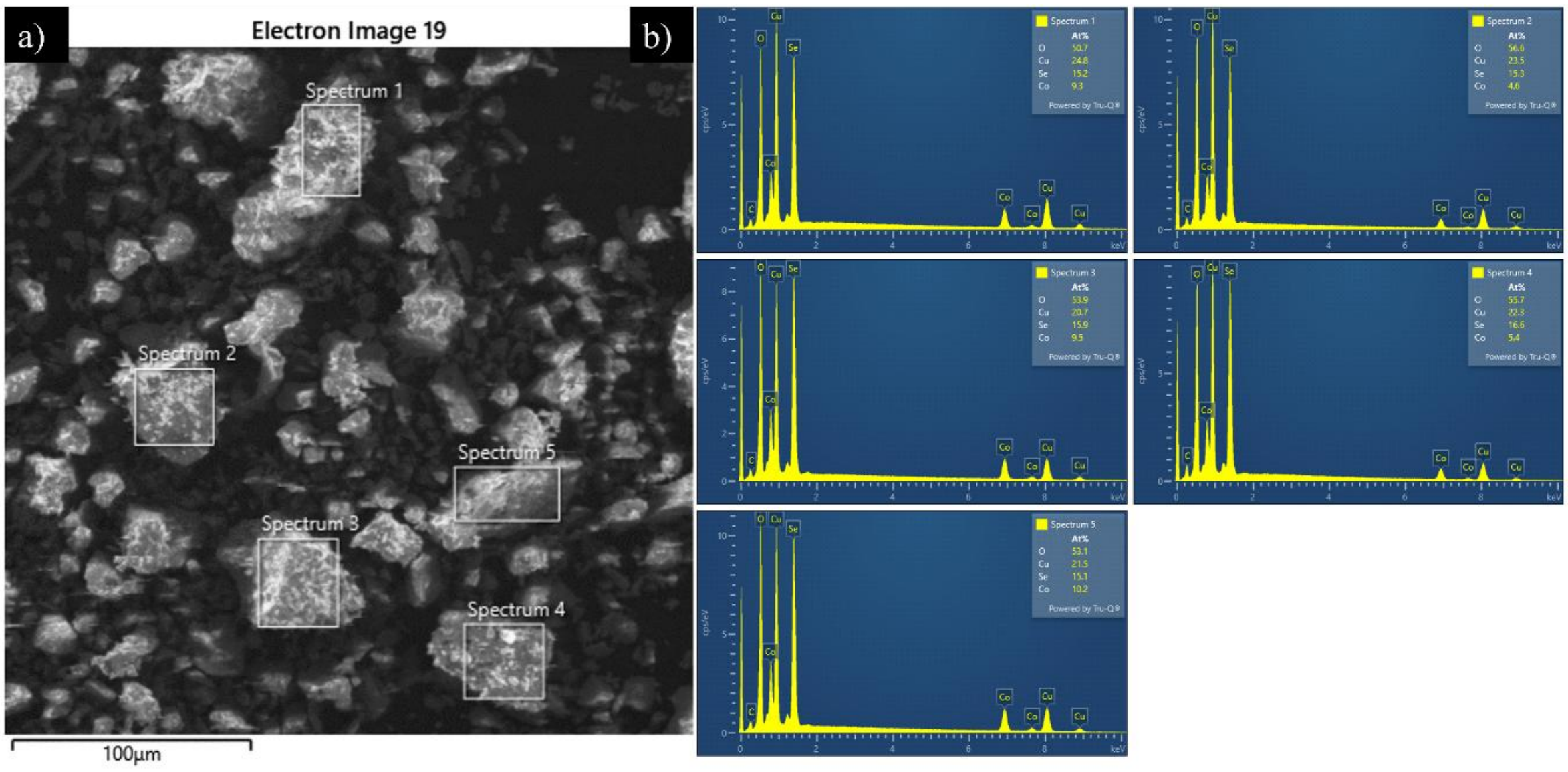

**FIG. S4.** a) A backscattered SEM image of the polycrystalline $(Cu_{0.8}Co_{0.2})_2OSeO_3$ sample with five white boxes identifying the regions that were scanned with EDS. The image was measured under a WD of 15.3 mm and a magnification of 954x. b) The five corresponding EDS spectra.

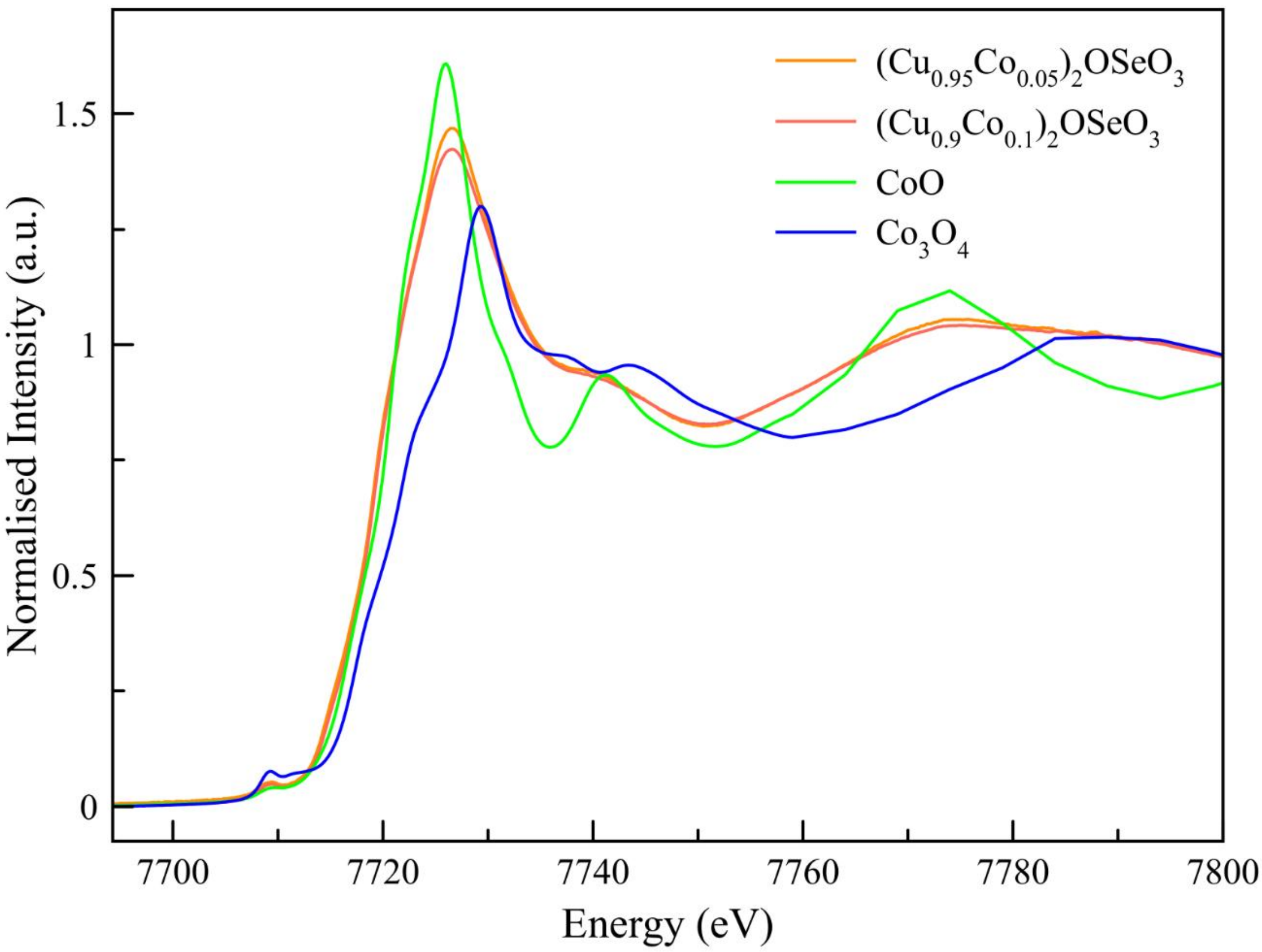

**FIG. S5.** The XANES spectra of the cobalt *K*-edge of $(Cu_{0.95}Co_{0.05})_2OSeO_3$ and $(Cu_{0.9}Co_{0.1})_2OSeO_3$ compared to reference standards CoO and $Co_3O_4$.

In order for cobalt to dope into either of the $Cu^{2+}$ sites in the crystal structure without impacting the overall stoichiometry of the compound, it must be in the +2 oxidation state. Using X-ray absorption spectroscopy (XAS), it was confirmed that cobalt in the structure was in the +2 oxidation state, as both the absorption edges from the measured samples align with the absorption edge of the CoO reference at 7726 eV, shown in Fig. S5. No shift to higher energy (signs of partial oxidation to $Co^{3+}$) can be observed. The $x = 0.05$ and $x = 0.1$ samples were compared against reference standards for $Co^{2+}$ and $Co^{3+}$ in CoO and $Co_3O_4$, respectively. Both spectra show no difference between the two, indicating the same doping mechanism in the crystal structure.

The occupancy of $Co^{2+}$ can be refined for the $x = 0.1$ and 0.2 samples, but not the $x = 0.02$ and 0.05 samples, as they did not have sufficient cobalt content to be resolved. Examining the occupancy for the $x = 0.1$ sample shows that the total doping is 3(1)% with the preference for $Co^{2+}$ inclusion into the Cu2 site, 4(1)%, over the Cu1 site, 0(2)%. Meanwhile, for the $x = 0.2$ sample, it shows that the total doping is 7(1)% with the preference for $Co^{2+}$ inclusion into the Cu1 site, 9(2)% over the Cu2 site, 6(1)%. The two copper site occupancies from the Rietveld refinement are shown in Table S2, along with the atomic positions of all the unique atoms in the unit cell. The refined Co-doping percentage differs from the calculated value from EDS, which is likely due to the low amount of total cobalt in the sample. Despite the difference in neutron scattering lengths, the refinement could not accurately resolve the two ions. However, since $Co^{2+}$ could be refined in the crystal structure of the $x = 0.1$ and 0.2 samples, and EDS measured the presence of cobalt in all doped samples, it can be inferred that all three samples have been Co-doped successfully in the crystal structure. As the actual doping is within the range of the nominal doping, the samples will be referred to by their nominal percentages.

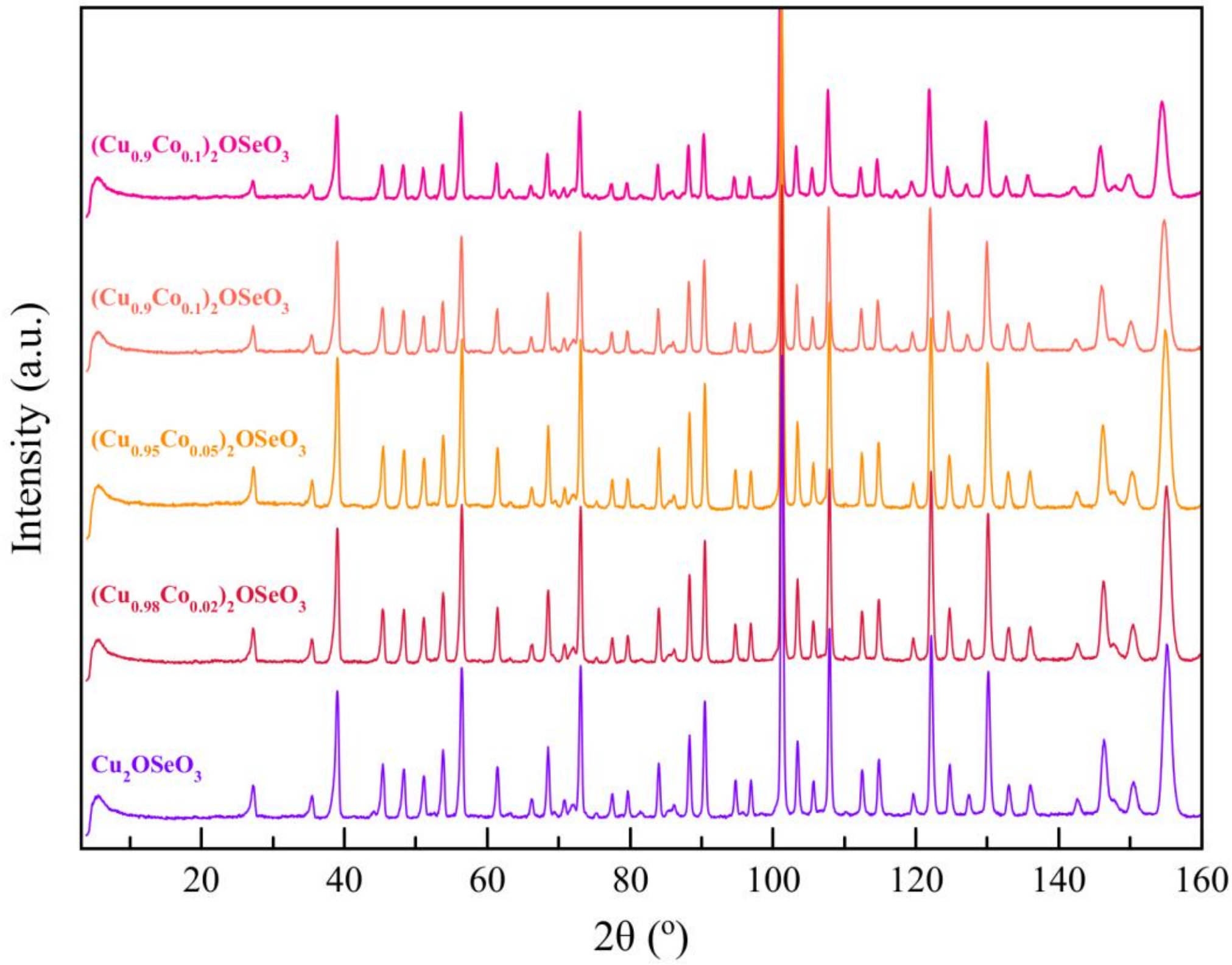

**FIG. S6.** Room temperature high-resolution NPD patterns of the polycrystalline $(Cu_{1-x}Co_x)_2OSeO_3$ ($0 \leq x \leq 0.2$) samples collected at 2.44 Å.

Fig. S6 shows the room temperature high-resolution NPD patterns of the polycrystalline $(Cu_{1-x}Co_x)_2OSeO_3$ ($0 \leq x \leq 0.2$) samples. Using the same Rietveld refinement method as the X-ray data, the same structural information can be extracted from NPD data and compared against, as summarised in Table S8, along with the Rietveld R-factors shown in Table S9. For the $x$ = 0.02 and 0.05 samples, the cobalt occupancy was fixed with the nominal percentage of $Co^{2+}$ inclusion for the refinements, while for the $x$ = 0.1 and 0.2 samples, the cobalt occupancy was able to be refined. Similarly to the refined pXRD data, it can be observed that $Co^{2+}$ has been successfully doped into the structure by an increase in the lattice parameters (lattice constants and volume), shown in Figure S9. The XRD and NPD data were refined against models taken from the Inorganic Crystal Structure Database (ICSD). The reference codes used were: $Cu_2OSeO_3$ (60652), $CuSeO_3$ (29507), CuO (16025), $Co_3O_4$ (36256), and $CoSeO_3$ (80401).

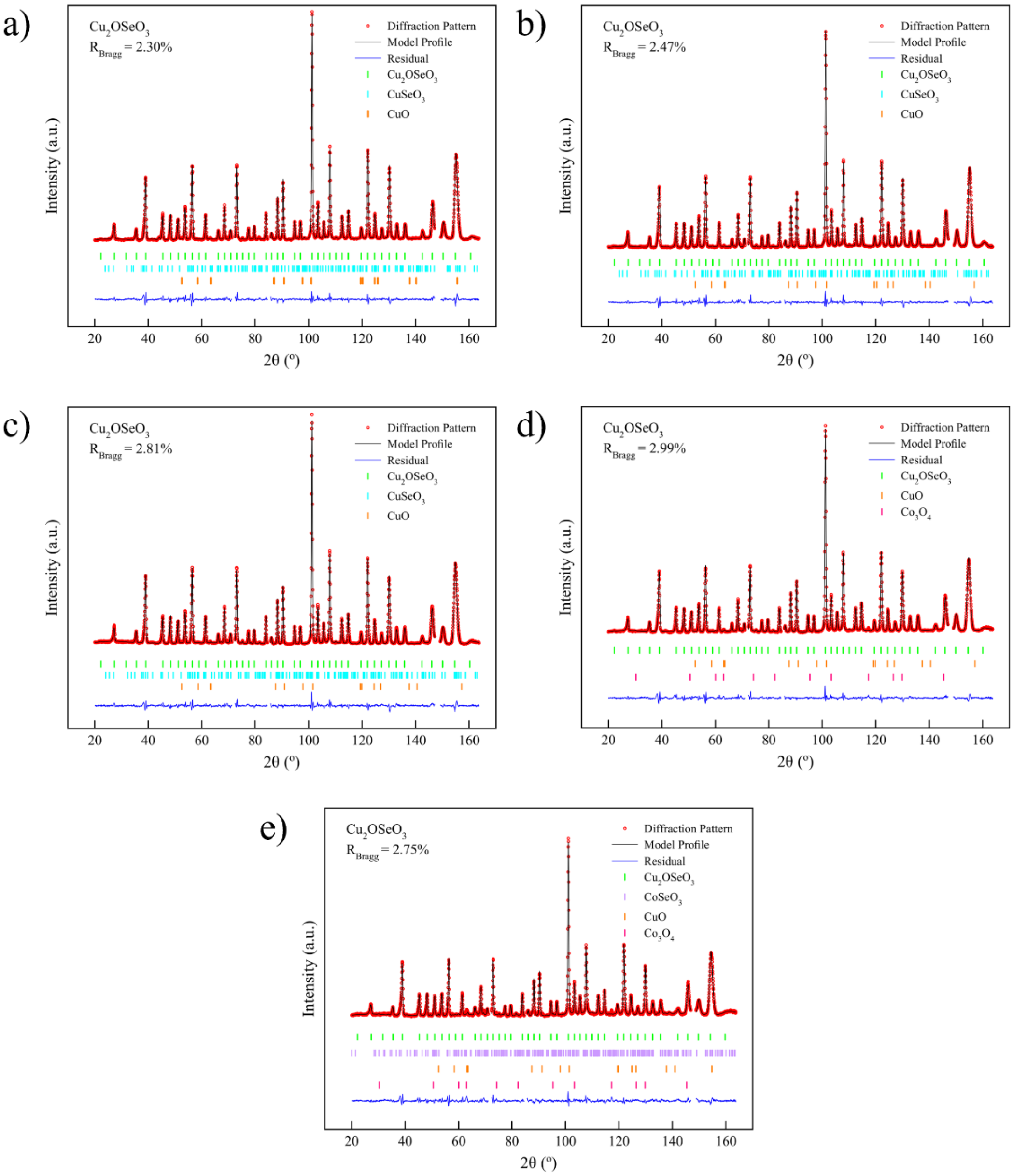

**FIG. S7.** NPD Rietveld Refinements. Rietveld refinements of room temperature high-resolution ECHIDNA NPD data of the polycrystalline $(Cu_{1-x}Co_x)_2OSeO_3$ samples: a) $Cu_2OSeO_3$, b) $(Cu_{0.98}Co_{0.02})_2OSeO_3$, c) $(Cu_{0.95}Co_{0.05})_2OSeO_3$, d) $(Cu_{0.9}Co_{0.1})_2OSeO_3$, and e) $(Cu_{0.8}Co_{0.2})_2OSeO_3$, collected at 2.44 Å.

The neutron refinements were carried out in a similar manner using the FullProf software, but due to stray instrumental diffraction, three regions were excluded from the Rietveld refinement at 2θ values: 71.23 – 72.46, 84.81 – 85.75 and 147.23 – 149.27°.

| Atom | *x* | *y* | *z* | $B_{iso}$ | Occupancy | Multiplicity |
|---|---|---|---|---|---|---|
| **Cu1** | 0.88467(53) | 0.88467(53) | 0.88467(53) | 1.037(56) | 0.3331(62) | 4 |
| **Co1** | 0.88467(53) | 0.88467(53) | 0.88467(53) | 1.037(56) | 0.0002(62) | 4 |
| **Cu2** | 0.13441(52) | 0.11934(42) | -0.12567(56) | 1.037(56) | 0.9562(96) | 12 |
| **Co2** | 0.13441(52) | 0.11934(42) | -0.12567(56) | 1.037(56) | 0.0438(96) | 12 |
| **O1** | 0.01256(46) | 0.01256(46) | 0.01256(46) | 0.993(51) | 0.3333 | 4 |
| **O2** | 0.75844(49) | 0.75844(49) | 0.75844(49) | 0.993(51) | 0.3333 | 4 |
| **Se1** | 0.45854(32) | 0.45854(32) | 0.45854(32) | 0.694(62) | 0.3333 | 4 |
| **O3** | 0.27043(42) | 0.48094(5)3 | 0.46814(47) | 0.993(51) | 1.0000 | 12 |
| **Se2** | 0.21171(35) | 0.21171(35) | 0.21171(35) | 0.694(62) | 0.3333 | 4 |
| **O4** | 0.27108(37) | 0.19024(56) | 0.03257(50) | 0.993(51) | 1.0000 | 12 |

**TABLE S2.** Atom parameters for the polycrystalline $(Cu_{0.9}Co_{0.1})_2OSeO_3$ sample. The atom parameters of the main phase, $Cu_2OSeO_3$, show the refined occupancy of cobalt in the copper sites, showing a preference for the Cu2 site from the NPD data.

| Atom | x | y | z | $B_{iso}$ | Occupancy | Multiplicity |
|---|---|---|---|---|---|---|
| **Cu1** | 0.88272(79) | 0.88272(79) | 0.88272(79) | 1.457(70) | 0.3024(77) | 4 |
| **Co1** | 0.88272(79) | 0.88272(79) | 0.88272(79) | 1.457(70) | 0.0309(77) | 4 |
| **Cu2** | 0.13520(73) | 0.11975(55) | -0.12850(75) | 1.457(70) | 0.9395(119) | 12 |
| **Co2** | 0.13520(73) | 0.11975(55) | -0.12850(75) | 1.457(70) | 0.0605(119) | 12 |
| **O1** | 0.01077(53) | 0.01077(53) | 0.01077(53) | 1.205(65) | 0.3333(0) | 4 |
| **O2** | 0.75869(56) | 0.75869(56) | 0.75869(46) | 1.205(65) | 0.3333(0) | 4 |
| **Se1** | 0.45843(40) | 0.45843(40) | 0.45843(40) | 1.140(78) | 0.3333(0) | 4 |
| **O3** | 0.26879(48) | 0.48044(66) | 0.46817(57) | 1.205(65) | 1.0000(0) | 12 |
| **Se2** | 0.21218(43) | 0.21218(43) | 0.21218(43) | 1.140(78) | 0.3333(0) | 4 |
| **O4** | 0.27183(45) | 0.18842(62) | 0.03419(61) | 1.205(65) | 1.0000(0) | 12 |

**TABLE S3.** Atom parameters for the polycrystalline $(Cu_{0.8}Co_{0.2})_2OSeO_3$ sample. The atom parameters of the main phase, $Cu_2OSeO_3$, show the refined occupancy of cobalt in the copper sites, showing a slight preference for the Cu1 site over the Cu2 site from the NPD data.

| Nominal Composition ($x$) | Phases | | | | | | | | | |
|---|---|---|---|---|---|---|---|---|---|---|
| | $Cu_2OSeO_3$ (%) | | $CuSeO_3$ (%) | | CuO (%) | | $Co_3O_4$ (%) | | $CoSeO_3$ (%) | |
| | *XRD* | *NPD* | *XRD* | *NPD* | *XRD* | *NPD* | *XRD* | *NPD* | *XRD* | *NPD* |
| **0** | 96.14(17) | 95.43(75) | 3.37(5) | 4.16(17) | 0.49(2) | 0.41(6) | n / a | | n / a | |
| **0.02** | 98.41(15) | 98.41(77) | 1.10(4) | 1.38(15) | 0.49(2) | 0.21(6) | n / a | | n / a | |
| **0.05** | 98.37(15) | 97.66(72) | 0.95(3) | 1.79(15) | 0.69(2) | 0.55(6) | n / a | | n / a | |
| **0.1** | 98.51(17) | 99.14(103) | n / a | | 0.90(3) | 0.67(7) | 0.59(2) | 0.19(3) | n / a | |
| **0.2** | 93.23(27) | 89.80(117) | n / a | | 1.64(5) | 2.37(36) | 1.27(5) | 0.22(3) | 3.86(6) | 7.61(37) |

**TABLE S4.** Phase composition for the corresponding polycrystalline $(Cu_{1-x}Co_x)_2OSeO_3$ ($0 \leq x \leq 0.2$) samples. Phase composition weight percentages based on the Rietveld refinements of the pXRD and NPD data.

All Co-doped samples have > 98% phase purity besides the $x = 0.2$ sample, with minor impurities of $CuSeO_3$ and CuO for samples $0 \leq x \leq 0.05$, while CuO and $Co_3O_4$ for the $x = 0.1$ sample, and lastly, CuO, $Co_3O_4$, and $CoSeO_3$ for the $x = 0.2$ sample (Table S4). Interestingly, there was an estimated 0.5 – 1% of an unidentifiable phase that was present in the $x = 0.1$ sample, which may share similarities with the Zn-doped phase that was found by Sukhanov *et al.*, as a probable dual transition metal-doped phase with a novel crystal structure.[23]

| $(Cu_{1-x}Co_x)_2OSeO_3$ | $x = 0$ | $x = 0.02$ | $x = 0.05$ | $x = 0.1$ | $x = 0.2$ |
|---|---|---|---|---|---|
| ***a*** **(Å)** | 8.92359(1) | 8.92574(1) | 8.9278(1) | 8.93086(1) | 8.93590(2) |
| ***V*** **(Å³)** | 710.589(1) | 711.104(1) | 711.595(1) | 712.327(2) | 713.533(3) |
| **Cu1-O1 (Å)** | 1.942(4) | 1.932(4) | 1.920(3) | 1.928(4) | 1.944(6) |
| **Cu1-O2 (Å)** | 1.943(4) | 1.933(4) | 1.907(4) | 1.945(4) | 1.881(7) |
| **Cu1-O3 (Å)** | 2.092(3) | 2.086(3) | 2.098(3) | 2.091(4) | 2.090(6) |
| **Cu2-O1 (Å)** | 1.920(4) | 1.923(3) | 1.927(3) | 1.926(4) | 1.922(6) |
| **Cu2-O2 (Å)** | 1.961(4) | 1.965(4) | 1.977(4) | 1.961(4) | 1.993(7) |
| **Cu2-O3 (Å)** | 2.019(3) | 2.030(3) | 2.010(3) | 2.009(4) | 2.018(6) |
| **Cu2-O4 (Å)** | 1.975(3) | 1.968(3) | 1.973(3) | 1.993(3) | 1.974(7) |
| **Cu2-O4* (Å)** | 2.293(3) | 2.305(3) | 2.294(3) | 2.287(3) | 2.294(5) |
| **O1-Cu1-O2 (deg.)** | 180.0(3) | 180.0(3) | 180.0(3) | 180.0(3) | 180.0(6) |
| **O1-Cu1-O3 (deg.)** | 77.9(2) | 77.98(18) | 77.30(18) | 77.3(2) | 77.1(4) |
| **O2-Cu1-O3 (deg.)** | 102.1(3) | 102.0(2) | 102.7(3) | 102.7(3) | 102.9(5) |
| **O3-Cu1-O3 (deg.)** | 115.7(3) | 115.8(2) | 115.3(2) | 115.3(3) | 115.2(5) |
| **O1-Cu2-O2 (deg.)** | 170.6(3) | 170.5(3) | 169.9(3) | 170.4(3) | 168.7(6) |
| **O1-Cu2-O3 (deg.)** | 80.2(2) | 79.55(18) | 79.3(2) | 79.4(2) | 79.4(4) |
| **O1-Cu2-O4 (deg.)** | 91.7(2) | 91.7(2) | 92.4(2) | 92.3(2) | 92.1(4) |
| **O1-Cu2-O4* (deg.)** | 110.5(3) | 110.6(2) | 110.7(2) | 110.3(3) | 111.4(5) |
| **O1-Cu2-O4# (deg.)** | 59.11(18) | 59.45(15) | 59.81(16) | 59.76(18) | 59.5(3) |
| **O2-Cu2-O3 (deg.)** | 101.1(3) | 101.1(2) | 100.5(2) | 101.4(3) | 100.3(5) |
| **O2-Cu2-O4 (deg.)** | 87.2(2) | 87.7(2) | 87.8(2) | 87.0(2) | 88.2(5) |
| **O2-Cu2-O4* (deg.)** | 78.9(2) | 78.9(2) | 79.4(2) | 79.3(2) | 79.8(4) |
| **O2-Cu2-O4# (deg.)** | 113.3(2) | 113.2(2) | 113.0(2) | 112.8(2) | 112.4(5) |
| **O3-Cu2-O4 (deg.)** | 171.7(3) | 171.1(3) | 171.7(3) | 171.6(3) | 171.5(6) |
| **O3-Cu2-O4* (deg.)** | 88.65(19) | 89.20(18) | 89.02(19) | 88.7(2) | 89.3(4) |
| **O3-Cu2-O4# (deg.)** | 121.6(2) | 121.2(2) | 121.4(2) | 121.7(2) | 121.2(4) |
| **O4-Cu2-O4 (deg.)** | 92.7(2) | 92.5(2) | 93.1(2) | 93.2(2) | 93.1(4) |
| **O4-Cu2-O4* (deg.)** | 53.92(15) | 53.96(15) | 53.65(15) | 53.37(15) | 53.6(3) |
| **O4-Cu2-O4# (deg.)** | 141.7(2) | 141.8(2) | 141.9(2) | 141.6(2) | 142.1(4) |

**TABLE S5.** The lattice parameters, bond lengths and bond angles for the polycrystalline $(Cu_{1-x}Co_x)_2OSeO_3$ ($0 \leq x \leq 0.2$) samples. All the data is refined from room temperature synchrotron pXRD data.

The undoped data, for comparison's sake, has been taken from the reference:

Ferguson, A. J., Vás, M., Vella, E., Pervez, M. d. F., Gilbert, E. P., Ulrich C., Yick, S., Söhnel, T. Skyrmion Stabilisation and Critical Behaviour in Tellurium-doped $Cu_2OSeO_3$. Commun Mater 6, 85, (2025). https://doi.org/10.1038/s43246-025-00804-4, with the permission of the authors.

| Nominal Composition ($x$) | Cu-Cu Distances (Å) | | | |
|---|---|---|---|---|
| | sAFM Cu1-Cu2 | wAFM Cu1-Cu2 | wFM Cu2-Cu2 | sFM Cu2-Cu2 |
| **0** | 3.0474(9) | 3.3098(10) | 3.0529(9) | 3.2230(10) |
| **0.02** | 3.0511(8) | 3.3093(9) | 3.0527(9) | 3.2234(8) |
| **0.05** | 3.0545(9) | 3.3089(9) | 3.0553(8) | 3.2215(10) |
| **0.1** | 3.0540(10) | 3.3108(10) | 3.0553(11) | 3.2247(11) |
| **0.2** | 3.0624(16) | 3.3125(16) | 3.0607(16) | 3.2181(17) |

**TABLE S6.** Cu-Cu distances for the polycrystalline $(Cu_{1-x}Co_x)_2OSeO_3$ ($0 \leq x \leq 0.2$) samples refined from room temperature synchrotron pXRD data. Cu1-Cu2 and Cu2-Cu2 distances are labelled with strong and weak ferromagnetic (sFM and wFM), and strong and weak antiferromagnetic (sAFM and wAFM).

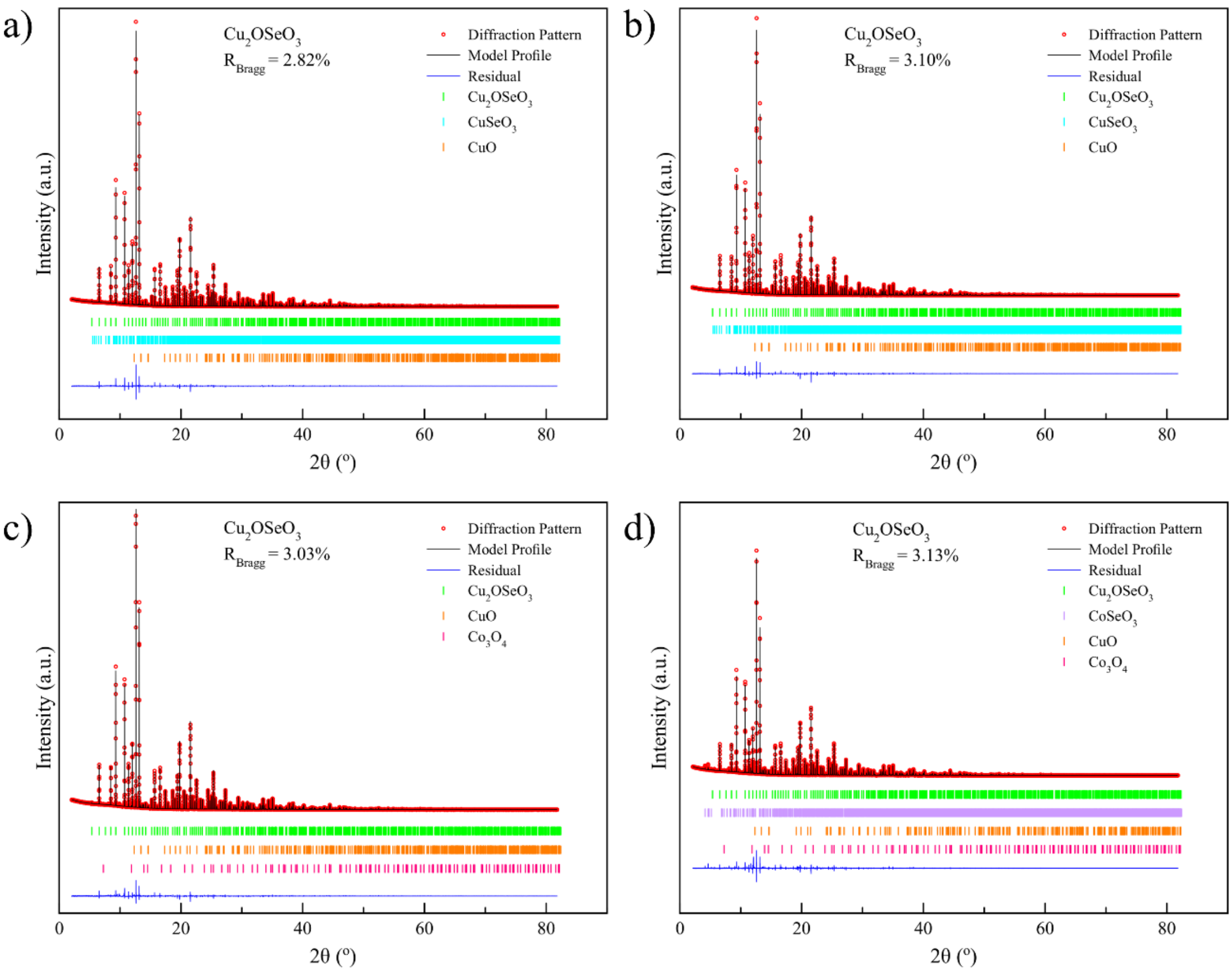

**FIG. S8.** The Rietveld refinements of the room temperature synchrotron pXRD data for the polycrystalline $(Cu_{1-x}Co_x)_2OSeO_3$ samples: a) $(Cu_{0.98}Co_{0.02})_2OSeO_3$, b) $(Cu_{0.95}Co_{0.05})_2OSeO_3$, c) $(Cu_{0.9}Co_{0.1})_2OSeO_3$, and d) $(Cu_{0.8}Co_{0.2})_2OSeO_3$, collected at 0.59053(1) Å.

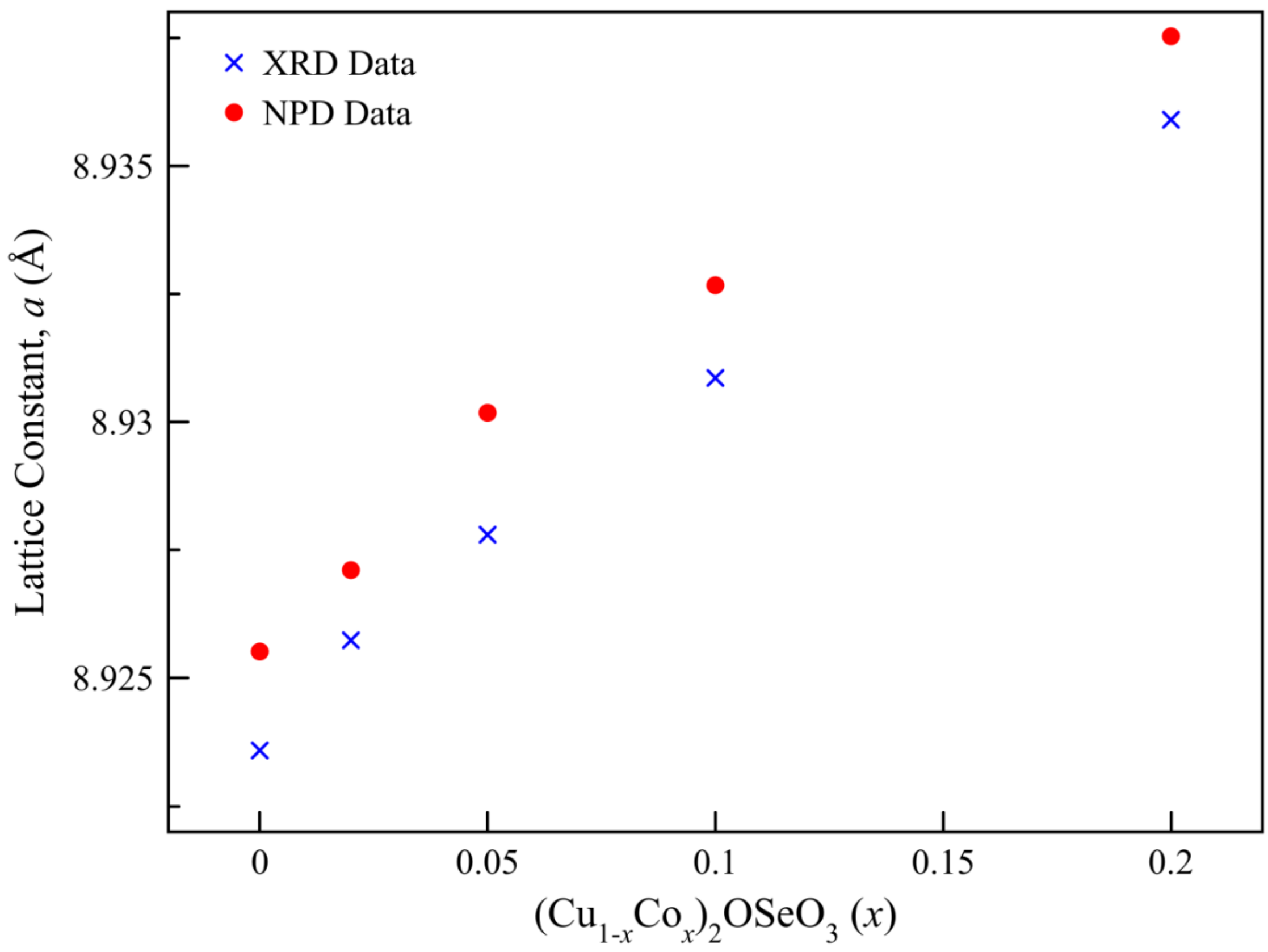

**FIG. S9.** Lattice constants for the corresponding polycrystalline $(Cu_{1-x}Co_x)_2OSeO_3$ $(0 \leq x \leq 0.2)$ samples refined from room temperature synchrotron pXRD and room temperature high-resolution NPD data. The error bars are excluded from the figure because they are too small to be visible.

| $(Cu_{1-x}Co_x)_2OSeO_3$ | $x = 0$ | | $x = 0.02$ | | $x = 0.05$ | | $x = 0.1$ | | $x = 0.2$ | |
|---|---|---|---|---|---|---|---|---|---|---|
| Technique | *XRD* | *NPD* | *XRD* | *NPD* | *XRD* | *NPD* | *XRD* | *NPD* | *XRD* | *NPD* |
| Cu1 Site Volume (Å$^3$) | 7.0381 | 7.148 | 6.9650 | 7.0758 | 6.9392 | 7.0195 | 6.9797 | 7.1177 | 6.8737 | 6.9789 |
| Cu2 Site Volume (Å$^3$) | 6.0184 | 5.9205 | 6.0737 | 5.9628 | 6.0794 | 5.9651 | 6.0381 | 5.9453 | 6.1570 | 6.0212 |

**TABLE S7.** The polyhedra volume for both Cu1 and Cu2 sites for the $(Cu_{1-x}Co_x)_2OSeO_3$ ($0 \leq x \leq 0.2$) samples refined from room temperature synchrotron pXRD data and room temperature NPD data.

| $(Cu_{1-x}Co_x)_2OSeO_3$ | $x = 0$ | $x = 0.02$ | $x = 0.05$ | $x = 0.1$ | $x = 0.2$ |
|---|---|---|---|---|---|
| ***a* (Å)** | 8.92552(3) | 8.92711(3) | 8.93018(2) | 8.93267(3) | 8.93753(4) |
| ***V* (Å$^3$)** | 711.050(4) | 711.430(4) | 712.165(3) | 712.761(4) | 713.925(5) |
| **Cu1-O1 (Å)** | 1.962(6) | 1.945(5) | 1.972(5) | 1.979(6) | 1.983(8) |
| **Cu1-O2 (Å)** | 1.932(6) | 1.945(5) | 1.932(5) | 1.954(6) | 1.920(9) |
| **Cu1-O3 (Å)** | 2.109(6) | 2.098(5) | 2.089(5) | 2.094(6) | 2.091(9) |
| **Cu2-O1 (Å)** | 1.905(5) | 1.887(5) | 1.901(5) | 1.902(6) | 1.932(8) |
| **Cu2-O2 (Å)** | 1.972(6) | 1.994(5) | 1.977(5) | 1.966(6) | 1.949(8) |
| **Cu2-O3 (Å)** | 2.030(5) | 2.017(5) | 2.039(5) | 2.040(6) | 2.026(8) |
| **Cu2-O4 (Å)** | 1.960(6) | 1.971(5) | 1.962(5) | 1.973(6) | 1.996(8) |
| **Cu2-O4* (Å)** | 2.263(6) | 2.276(6) | 2.272(6) | 2.268(6) | 2.274(8) |
| **O1-Cu1-O2 (deg.)** | 180.0(4) | 180.0(4) | 179.9(4) | 180.0(5) | 180.0(5) |
| **O1-Cu1-O3 (deg.)** | 77.5(3) | 77.7(3) | 77.3(3) | 77.4(3) | 76.4(4) |
| **O2-Cu1-O3 (deg.)** | 102.5(4) | 102.3(3) | 102.7(3) | 102.6(4) | 103.6(5) |
| **O3-Cu1-O3 (deg.)** | 115.4(4) | 115.6(3) | 115.3(3) | 115.4(4) | 114.6(5) |
| **O1-Cu2-O2 (deg.)** | 170.9(4) | 170.5(4) | 170.4(4) | 170.7(5) | 170.4(5) |
| **O1-Cu2-O3 (deg.)** | 80.7(3) | 81.1(3) | 80.1(3) | 80.5(4) | 79.1(4) |
| **O1-Cu2-O4 (deg.)** | 93.6(3) | 93.6(3) | 93.3(3) | 92.8(4) | 92.1(4) |
| **O1-Cu2-O4* (deg.)** | 110.8(3) | 111.3(3) | 111.3(3) | 111.1(4) | 110.3(4) |
| **O1-Cu2-O4$^\#$ (deg.)** | 59.6(2) | 59.7(2) | 59.4(2) | 59.1(3) | 59.5(3) |
| **O2-Cu2-O3 (deg.)** | 99.9(3) | 99.6(3) | 100.5(3) | 100.8(4) | 102.1(5) |
| **O2-Cu2-O4 (deg.)** | 86.1(3) | 85.8(3) | 86.2(3) | 85.9(4) | 86.6(5) |
| **O2-Cu2-O4* (deg.)** | 78.3(3) | 78.1(3) | 78.3(3) | 78.3(3) | 79.3(4) |
| **O2-Cu2-O4$^\#$ (deg.)** | 113.4(3) | 113.0(3) | 113.1(3) | 113.3(3) | 112.8(4) |
| **O3-Cu2-O4 (deg.)** | 174.0(4) | 174.5(4) | 173.3(4) | 173.3(4) | 171.2(5) |
| **O3-Cu2-O4* (deg.)** | 88.6(3) | 88.5(3) | 89.2(3) | 88.7(3) | 89.8(4) |
| **O3-Cu2-O4$^\#$ (deg.)** | 121.8(3) | 122.3(3) | 121.0(3) | 121.3(3) | 120.6(4) |
| **O4-Cu2-O4 (deg.)** | 91.6(3) | 92.0(3) | 92.5(3) | 93.2(4) | 92.7(4) |
| **O4-Cu2-O4* (deg.)** | 55.8(2) | 55.2(2) | 55.0(2) | 54.5(3) | 53.8(3) |
| **O4-Cu2-O4$^\#$ (deg.)** | 142.4(3) | 142.5(3) | 142.7(3) | 142.5(3) | 141.5(4) |

**TABLE S8.** The lattice parameters, bond lengths and bond angles for the polycrystalline $(Cu_{1-x}Co_x)_2OSeO_3$ ($0 \leq x \leq 0.2$) samples. All the data is refined from room temperature high-resolution NPD data.

| Nominal Composition ($x$) | $R_{Bragg}$ Factor | | $R_f$ Factor | | $Chi^2$ | |
|---|---|---|---|---|---|---|
| | *XRD* | *NPD* | *XRD* | *NPD* | *XRD* | *NPD* |
| **0** | 2.77 | 2.29 | 2.86 | 1.59 | 6.98 | 4.86 |
| **0.02** | 2.82 | 2.47 | 2.45 | 1.62 | 4.32 | 5.22 |
| **0.05** | 3.11 | 2.81 | 3.03 | 1.81 | 4.41 | 4.02 |
| **0.1** | 3.03 | 2.99 | 3.82 | 2.21 | 5.26 | 4.60 |
| **0.2** | 3.13 | 2.75 | 2.66 | 2.04 | 15.1 | 4.41 |

**TABLE S9.** The Rietveld R-factors ($R_{Bragg}$, $R_f$) and $Chi^2$ from the XRD and NPD Rietveld refinements for corresponding polycrystalline $(Cu_{1-x}Co_x)_2OSeO_3$ ($0 \leq x \leq 0.2$) samples.

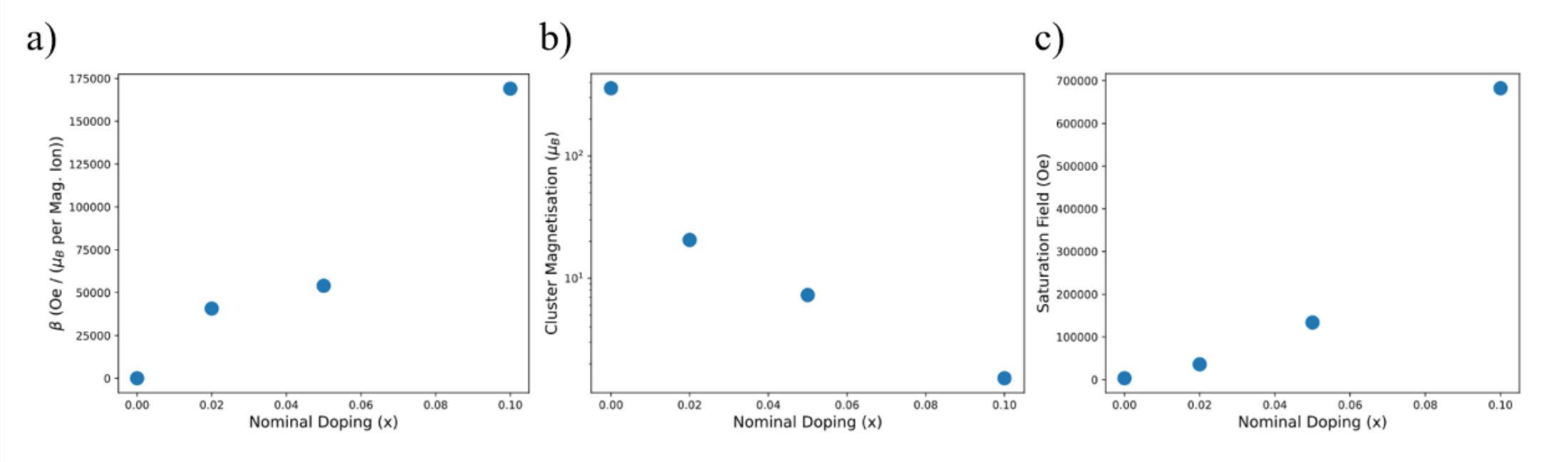

**FIG. S10.** Fitting parameters from the anhysteretic extended Langevin model against nominal doping for the polycrystalline $(Cu_{1-x}Co_x)_2OSeO_3$ ($0 \leq x \leq 0.1$) samples. a) The inter-cluster interaction parameter, $\beta$. b) the magnetisation of individual clusters. c) The saturation field, as defined by the condition $M(H_{sat})=0.975M_{sat}$.

The skyrmion peak intensity of the Co-doped samples has a similar peak intensity to the undoped sample, but the maximum intensity is shifted to lower temperatures. The peak maxima decrease to 53 K in the $x$ = 0.02 sample and then down to 47 K in the $x$ = 0.05 sample. The skyrmion temperature stability has also significantly increased, especially for the $x$ = 0.05 sample, which has skyrmions appearing from 4 K. The $x$ = 0.02 sample has a temperature range of 45 – 56 K, while the $x$ = 0.05 sample has a temperature range of 4 – 60 K. Comparing to the magnetisation data, it is likely that the extended temperature range is due to the effects of a remanent field inside the magnet resulting in an unintentional field cooling protocol. This has been previously reported during field-cooling measurements of the undoped sample, where the high-temperature skyrmion phase has been observed as metastable skyrmions throughout the whole temperature range.[45] It was intended that all the measurements be carried out after ZFC; however, for the $x$ = 0.05 sample, the decrease in the critical field for the formation of the skyrmion phase allows for the formation of metastable skyrmions whilst cooling down with the remanent field.[45]

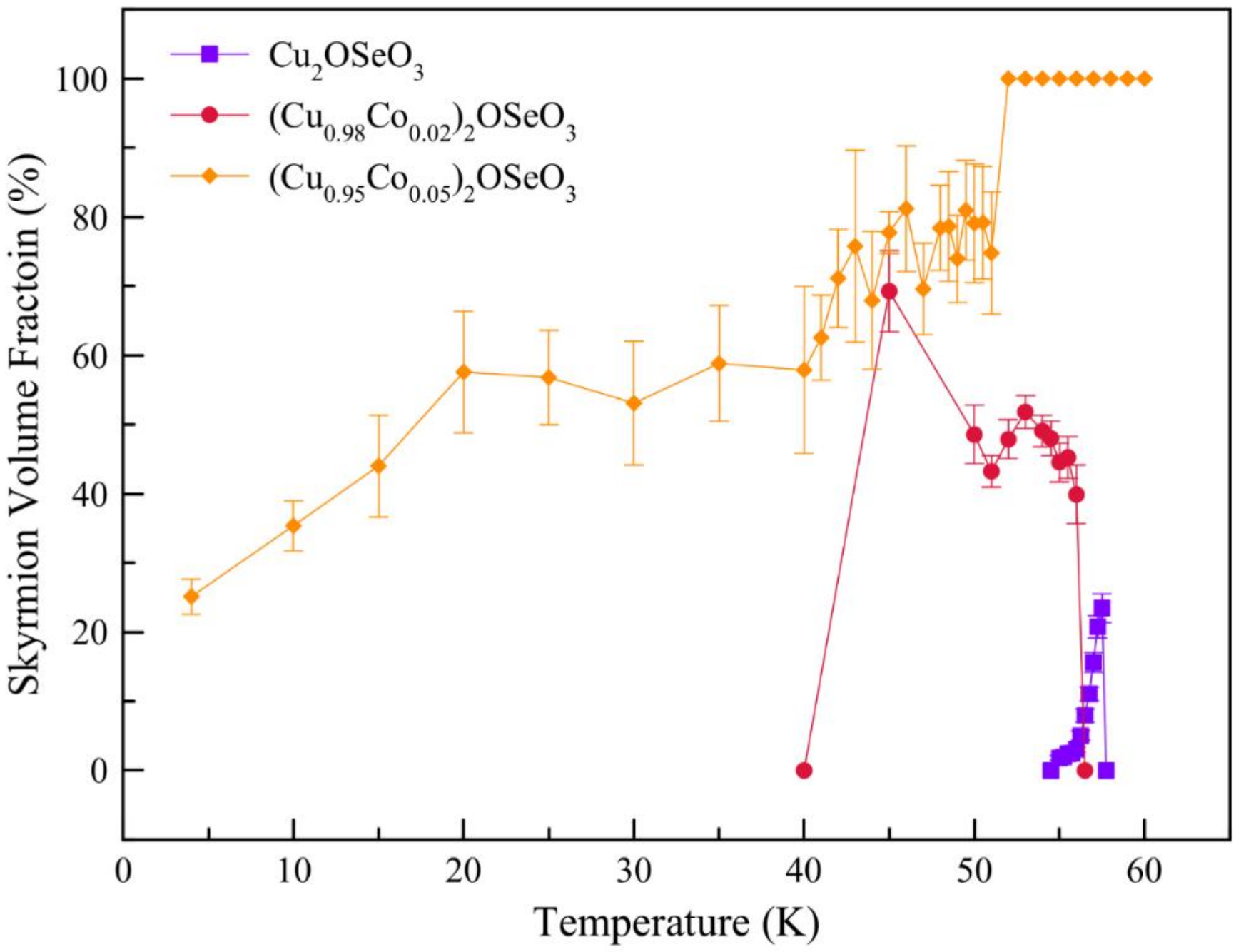

**FIG. S11.** The skyrmion volume fraction percentage determined from the SANS peak intensities of the skyrmion and conical phase as a function of temperature at 200 Oe for the $Cu_2OSeO_3$ sample and 250 Oe for the $(Cu_{0.98}Co_{0.02})_2OSeO_3$ and $(Cu_{0.95}Co_{0.05})_2OSeO_3$ samples.